\documentclass[aps,prx,twocolumn,preprintnumbers,superscriptaddress,amsmath]{revtex4}
\usepackage{graphicx}
\usepackage{dcolumn}
\usepackage{bm}
\usepackage{soul}
\usepackage{color}
\usepackage{xcolor}
\usepackage{amsmath}
\usepackage{amssymb}
\usepackage{amsmath}
\usepackage{float}
\usepackage{natmove}
\usepackage{multirow}
\usepackage{setspace}
\usepackage{array}
\usepackage{booktabs}
\usepackage{rotating}
\usepackage{ulem}
\usepackage{tabularx}
\usepackage[colorlinks=true,linkcolor=blue,citecolor=blue,urlcolor=blue]{hyperref}
\usepackage{xr-hyper}
\usepackage{}
\usepackage{physics}
\usepackage{mathrsfs}
\usepackage{verbatim} 

\newcommand{\yambo}{\textsc{yambo}}

\definecolor{Grey}{rgb}{0.50,0.50,0.50}
\definecolor{Blu}{rgb}{0.00,0.00,1.00}
\definecolor{Red}{rgb}{1.00,0.00,0.00}
\definecolor{Green}{rgb}{0.00,0.60,0.00}
\definecolor{Magenta}{rgb}{0.60,0.00,0.60}
\definecolor{BluBondi}{rgb}{0.00,0.58,0.71}
\definecolor{Orange}{rgb}{0.95,0.46,0.17}
\definecolor{Red}{rgb}{1.00,0.00,0.00}

\newcommand{\editor}[2]{%
  \expandafter\newcommand\csname #1note\endcsname[1]{%
    \textcolor{#2}{(\textbf{#1:} \textit{##1})}}%
  \expandafter\newcommand\csname #1\endcsname[1]{%
    \textcolor{#2}{##1}}%
  \expandafter\newcommand\csname #1cancel\endcsname[1]{%
    \textcolor{#2}{\sout{##1}}}%
  \expandafter\newcommand\csname #1change\endcsname[2]{%
    \textcolor{#2}{\sout{##1} ##2}}%
  \newenvironment{#1text}{\color{#2}}{\color{black}}
}

\editor{MB}{orange}
\editor{EM}{green}
\editor{AR}{Red}
\editor{AF}{BluBondi}
\editor{DV}{Red}
\editor{CC}{Magenta}
\editor{DAL}{purple}

\newcommand{\suppinfo}{Supplemental Material~\cite{supp-info}}
\newcommand{\drop}[1]{}

\newcommand{\kk}{\mathbf{k}}
\newcommand{\q}{\mathbf{q}}

\newcommand{\G}{\mathbf{G}}

\renewcommand{\o}{\omega}
\renewcommand{\O}{\Omega}
\usepackage{adjustbox}

\begin{document}

\title{Benchmarking the plasmon-pole and multipole approximations in the Yambo Code using the GW100 dataset}

\author{M. Bonacci}
\affiliation{PSI Center for Scientific Computing, Theory and Data, 5232 Villigen PSI, Switzerland}
\affiliation{
National Centre for Computational Design and Discovery of Novel Materials (MARVEL), 5232 Villigen PSI, Switzerland
}
\affiliation{
 S3 Centre, Istituto Nanoscienze, CNR, 41125 Modena, Italy
}
\author{D. A. Leon}
\affiliation{
 Department of Mechanical Engineering and Technology Management, \\ Norwegian University of Life Sciences, NO-1432 Ås, Norway
}
\author{N. Spallanzani}
\affiliation{
 S3 Centre, Istituto Nanoscienze, CNR, 41125 Modena, Italy
}
\author{E. Molinari}
\affiliation{
 S3 Centre, Istituto Nanoscienze, CNR, 41125 Modena, Italy
}
\affiliation{Dipartimento di Scienze Fisiche, Informatiche e Matematiche, Universit$\grave{a}$ di Modena e Reggio Emilia, I-41125 Modena, Italy}
\author{D. Varsano}
\affiliation{
 S3 Centre, Istituto Nanoscienze, CNR, 41125 Modena, Italy
}
\author{A. Ferretti}
\affiliation{
 S3 Centre, Istituto Nanoscienze, CNR, 41125 Modena, Italy
}
\author{C. Cardoso}
\email[corresponding author:]{claudiamaria.cardosopereira@nano.cnr.it}
\affiliation{
 S3 Centre, Istituto Nanoscienze, CNR, 41125 Modena, Italy
}

%

\begin{abstract}
Verification and validation of electronic structure codes are essential to ensure reliable and reproducible results in computational materials science. While density functional theory has been extensively benchmarked, systematic assessments of many-body perturbation theory methods such as the GW approximation have only recently emerged, most notably through the GW100 dataset. In this work, we assess the numerical accuracy and convergence behavior of the GW implementation in the \yambo{} code using both the Godby–Needs plasmon-pole model and the recently introduced multipole approximation. Quasiparticle energies are compared against GW100 reference data to evaluate the performance, numerical stability, and consistency of these approaches.
\end{abstract}

\maketitle

\section{Introduction}
First-principles computational methods have become well-established in the realm of electronic structure and materials science, serving as invaluable tools that offer critical insight and steer experimental investigations, as well as playing a pivotal role in materials discovery and  characterization~\cite{Marzari2021NatMater,qian2014quantum, mounet2018two, fischer2006predicting,curtarolo_high-throughput_2013,hautier2012computer}. 
Once a theoretical framework is defined, such as density functional theory (DFT) or the GW approximation within many-body perturbation theory (MBPT), different numerical implementations should ideally yield equivalent results.
In practice, however, algorithmic choices, numerical parameters, and approximations can introduce discrepancies.
For this reason, verification (ensuring consistency across codes) and validation (assessing accuracy against experiments) have become key aspects of methodological development. Together, verification and validation (V\&V) are essential to establish the reliability, reproducibility, and predictive power of computational techniques.

During the last decade, several systematic comparisons have demonstrated that state-of-the-art DFT codes yield highly consistent results for the ground-state properties of solid-state systems~\cite{lejaeghere_error_2014, lejaeghere_reproducibility_2016, bosoni2023verify}. 
On the other hand, excited-state properties, such as quasi-particle energy levels of materials calculated with GW in MBPT, have been primarily compared with experiments~\cite{onida_electronic_2002,Faleev_2004,Schilfgaarde_self_2006,Shishkin_2007,Jiang_2016,liao2011testing,Barker2022PhysRevB}, while code-to-code V\&V studies 
have emerged more recently~\cite{rangel_repr_2020,Ren2021PhysicalReviewMaterials,Azizi2025ComputMaterSci,grobmann2025}. 
For example, the work by Rangel et al.~\cite{rangel_repr_2020} compares  G$_0$W$_0$ quasiparticles of selected solids (Si, Au, TiO$_2$, and ZnO), obtained with three different plane-wave (PW) codes (\yambo~\cite{yambo1,yambo2}, \textsc{BerkeleyGW}~\cite{BGW} (BGW) and \textsc{Abinit}~\cite{Romero2020}).
The comparison shows that, when equivalent convergence parameters, numerical schemes, and pseudopotentials are employed, quasiparticle energies agree within 100 meV. Their systematic analysis traced previously reported discrepancies to differences in the treatment of the Coulomb divergence, and frequency-integration schemes. By making these aspects consistent across codes, they achieved consistent results even for challenging systems such as rutile TiO$_2$ and ZnO~\cite{Louie2010_ZnO,stankovski_g_2011,Friedrich2011PRB}. 

Concerning molecules, early studies have focused on benchmarking different flavors of the GW approximation, including different DFT starting points, with respect to small data sets of experimental results and accurate quantum chemistry calculations based, e.g.,  on coupled cluster methods~\cite{Thygesen2010, Blase_Attaccalite, bruneval_starting_point_mol,kaplan_2016,li2022benchmark,marom2012benchmark}. 
A significant community effort was made to V\&V different GW implementations, using the larger GW100 dataset, which includes 100 different molecules~\cite{gw100,caruso_benchmark_2016,maggio+gwsmall, gov+west+gw100,vlcek2017stochastic,forster2021gw100,gw100_repo}. 
This effort is primarily focused on the single-shot G$_0$W$_0$ calculation of the vertical ionization potential (IP) and electron affinity (EA)~\cite{gw100, maggio+gwsmall, gov+west+gw100}, although results from self-consistent GW schemes have also been reported~\cite{caruso_benchmark_2016}. The chosen data set comprises only closed-shell molecules, thereby avoiding the well-known issues presented by open-shell systems~\cite{stanton_discussion_2003}.

In terms of both single eigenvalues and statistical deviations (e.g. mean absolute errors), the reported differences between codes using PW and localized basis sets are of the order of 200 meV. The discrepancies are mainly due to ($i$) the size of the adopted basis sets and to ($ii$) the frequency-dependence treatment of the screened Coulomb potential $W$ and of the self-energy $\Sigma$.
Concerning the frequency dependence, the performance of the different plasmon pole (PPA) models has been addressed in the literature for a few bulk materials~\cite{stankovski_g_2011,Miglio2012EPJB,Larson2013PRB}. For the GW100 dataset, however, only the Hybertsen and Louie (HL) model~\cite{hybertsen_electron_1986} has been benchmarked against full-frequency (FF) results, while a benchmark for the largely adopted Godby and Needs (GN) model~\cite{godby_metal-insulator_1989} is still lacking.

In this work, we provide the results for IP and EA of all 100 molecules of the GW100 set as computed within the plane-wave code \textsc{Yambo}, using two different frequency dependence treatments: the Godby-Needs plasmon pole approximation~\cite{godby_metal-insulator_1989} (GN-PPA) and the recently developed multipole approximation (MPA)~\cite{Valido_2021,leon2023efficient}.
Our results show that GN-PPA, at variance with HL-PPA, is in good agreement with GW FF data obtained from other PW codes, with an average deviation of only 190 meV. MPA further reduces the deviation to 143 meV, which is comparable to the relative deviation among other FF approaches.

The paper is organized as follows. In Sec.~\ref{techs} we provide details about 
the GW100 set, the adopted GW methodology and implementation, the numerical parameters used in the simulations, and the techniques employed to automate the calculations. In Sec.~\ref{sec:results} we present and discuss the IP and EA results for the  GW100 dataset, obtained with both GN-PPA and MPA. 
Our conclusions are drawn in Sec.~\ref{sec:conclusions}.

\section{Computational Approach}
\label{techs}
%
\subsection{The GW100 dataset}
%
The GW100 set comprises a collection of 100 closed-shell molecules~\cite{gw100}, selected to include a wide range of ionization potential (IP) energies (4--25 eV) as well as a variety of chemical bond arrangements. The set includes carbon-based covalently bonded compounds like C$_2$H$_2$, C$_2$H$_4$, C$_2$H$_6$, as well as ionic bonded molecules such as the alkaline metal halide LiF. There are also molecules with metal atoms, such as Ag$_2$, Li$_2$, K$_2$, or small metal clusters (Na$_4$, Na$_6$), and common molecules such as water and carbon mono- and di-oxide. 

All molecular structures considered in this study have been taken from the official GW100 repository~\cite{gw100_repo}, derived either from experiments, or optimized using DFT-PBE with the def2-QZVP basis set~\cite{gw100}. For the cases of CH$_2$CHBr and C$_6$H$_5$OH, for which there is more than one reported structure, we have considered the most recent ones, as described in Ref.~\cite{maggio+gwsmall}.
Following previous GW100 benchmarks, the present  calculations employ a single-shot $G_0W_0$ scheme based on KS-DFT eigenvalues and eigenvectors obtained with the PBE functional~\cite{Perdew-Burke-Ernzerhof1996PRL}, hereafter denoted as $G_0W_0$@PBE. 
This choice allows us to bypass the well-known dependence of $G_0W_0$ results on the starting point~\cite{holm_cancellation_2004,rinke_combining_2005,atalla_hybrid_2013}, and to focus the discussion on the different GW  implementations and underlying numerical approximations.

\subsection{DFT setup}
\label{sec:setup}
%
Even with the same exchange–correlation functional and molecular geometries, the outcome of $G_0W_0$ calculations depends on several technical aspects of the underlying DFT setup.  
These include the adopted basis set, the treatment of core electrons (e.g., via pseudopotentials), the definition of the simulation cell and boundary conditions, and the treatment of the long-range Coulomb interaction between periodic replicas.
In Table~\ref{tab:all_codes}, we summarize the methodologies and approximations implemented in the codes benchmarked with the GW100 dataset~\cite{gw100,gov+west+gw100}, as well as those implemented in \yambo{} and used in the present work (see details below).
%

\newcommand{\citel}[1]{\!$^{\text{\footnotesize\citenum{#1}}}$}
\begin{table*}
\begin{adjustbox}{max width=\linewidth}
\begin{tabular}{lccccccc}
\hline\hline
    &\\[-5pt]
    \textbf{Code} & {\bf basis set} & {\bf PP} & {\bf empty states} & {\bf $\chi$-basis} & {\bf $\Sigma_c$-integration} & {\bf QP solutions} & {\bf Refs.}\\[4pt]
    \hline\\[-5pt]
    TM~\citel{TM} & def2-QZVP & AE & all & LO & FA~\citel{van_setten_gw_2013} & largest weight & \citenum{gw100}\\[4pt]
    FHI-aims~\citel{blum_ab_2009} & def2-QZVP$^*$ & AE & all & LO & AC~\citel{ren_resolution-identity_2012,blum_ab_2009} & iterative & \citenum{gw100}\\[4pt]
    BGW~\citel{BGW} & PW & TM NC~\citel{TMNC} & truncated~\citel{deslippe_coulomb-hole_2013} & PW & HL-PPA~\citel{hybertsen_electron_1986}/FF~\citel{liu_numerical_2015} & lin./graphical & \citenum{gw100}\\[4pt]

    VASP~\citel{VASP} & PAW* & PAW & truncated & PW* & AC~\citel{kaltak_low_2014} & linearized & \citenum{maggio+gwsmall}\\[4pt]

    WEST~\citel{WEST} & PW & ONCV & none~\citel{umari_gw_2010,WEST} & PDEP*~\citel{wilson_iterative_2009} & CD~\citel{lebegue_implementation_2003} & secant/lin. & \citenum{gov+west+gw100}\\[4pt]

\large{\yambo}~\citel{yambo1,yambo2} & PW & ONCV & truncated~\citel{BG} & PW & GN-PPA~\citel{godby_metal-insulator_1989}  / MPA  ~\citel{Valido_2021} & linearized & this work\\[4pt]
\hline\hline
\end{tabular}
\end{adjustbox}
\caption{Basis set,  
approximations used in previous GW100 benchmarks and in the present work. The codes TM and FHI-aims make use of the resolution-of-identity (RI) technique to compute the four-center Coulomb integrals~\cite{gw100}. In the column \textit{basis set}, ${*}$ indicates results obtained after extrapolation. 
In the column {\textit{PP}}, we list the approach used for core and valence electrons: all electron (AE), Troullier–Martins norm conserving pseudopotentials (TM NC), 
projector augmented waves (PAW), optimized norm-conserving Vanderbilt pseudopotentials (ONCV).  In the column \textit{$\Sigma_c$-integration}, the acronyms refer to the frequency integration methods used for the self-energy: fully-analytic (FA), analytic continuation (AC) and contour deformation (CD), Godby-Needs plasmon pole model (GN-PPA), Multipole approximation (MPA).}
\label{tab:all_codes}
\end{table*}

The GW100 data were originally obtained using a variety of basis sets, including projector augmented waves (PAW)~\cite{arnaud_all-electron_2000,shishkin_implementation_2006}, linear augmented plane waves (LAPW)~\cite{kutepov_electronic_2012}, linearized muffin-tin orbitals (LMTO)~\cite{kotani_all-electron_2002}, and local orbital (LO) basis sets.
For a detailed comparison of ionization potentials (IPs) and electron affinities (EAs) obtained using DFT-PBE with different basis sets, we refer the reader to Ref.~\cite{gov+west+gw100}. 
In this work, the Kohn–Sham (KS) wave functions and charge densities are expanded in plane waves (PWs) under periodic boundary conditions (PBCs), as implemented in the \textsc{Quantum ESPRESSO} simulation package~\cite{2009qe,2017qe}. 
Each molecule is placed in a supercell with sufficient vacuum to suppress spurious interactions between replicas. 
To reduce the required vacuum spacing and the associated computational cost, the Martyna-Tuckerman method~\cite{martyna_t} is adopted to correct the error
introduced by the periodic replica of the system.

The treatment of core and valence electrons varies across different codes.  
Approaches based on pseudopotentials (PPs) reduce the number of explicitly treated electrons, simplifying the calculations while retaining accuracy for valence states.  
Nevertheless, different PP schemes (ultrasoft vs. norm-conserving), the use of nonlinear core corrections (NLCC), or the inclusion of semicore states in valence can significantly affect GW quasiparticle energies~\cite{rohlfing_quasiparticle_1995,rinke_combining_2005,li_impact_2012}.  
Here, we employ optimized norm-conserving Vanderbilt (ONCV) pseudopotentials from the SG15 library~\cite{oncv1,oncv2}, without NLCC~\cite{NLCC}.  
Details of the electronic configurations, including the semicore states treated as valence, are provided in the Supplementary Material~\cite{oncv2}.

Optimizing the supercell geometry is also essential for both numerical accuracy and computational efficiency of our simulations.  
Previous GW100 studies~\cite{gov+west+gw100} used simple cubic (SC) supercells with lattice parameters up to \textit{a}=25~\AA.  
In the present work, we adopt a face-centered cubic (FCC) geometry with lattice parameter \textit{a}=13~\AA, which reduces the supercell volume by a factor of $\sqrt{2}$ relative to the corresponding SC configuration. Tests performed on two molecular systems, with 10 and 56 atoms (the latter being the largest molecule in the GW100 set), show a residual IP/EA error of  $\sim$25 and $\sim$180~meV, respectively (see Sec.~II of the \suppinfo).

At the GW level, a real-space truncation of the Coulomb interaction is applied to both the screening and self-energy terms~\cite{rozzi_exact_2006}.  
This spherical Coulomb cutoff, with a diameter of approximately 12~\AA\ (i.e. 1~\AA\ smaller than the lattice parameter) analytically removes the divergence of the long-range Coulomb interaction $v(\q\to0)$ and ensures rapid convergence with respect to the size of the supercell.

\subsection{The quasiparticle equation}
%
Within the MBPT framework, quasiparticle (QP) energies are usually obtained perturbatively from DFT, as the solutions of the QP equation:
\begin{equation}
\varepsilon_{n\kk}^{\mathrm{QP}} =  
\varepsilon_{n\kk} + 
\mel{n\kk}{\Sigma(\varepsilon_{n\kk}^{\mathrm{QP}}) - v_{xc}}{n\kk},
\label{Eq_QP_full}
\end{equation}
where $\Sigma$ is the electron self-energy, $\{\varepsilon_{n\kk}, \ket{n\kk}\}$ are the reference Kohn–Sham (KS) eigenvalues and wavefunctions,  and $v_{xc}$ is the corresponding exchange–correlation potential.  
Equation~\eqref{Eq_QP_full} 
originates from
a nonlinear eigenvalue problem~\cite{Guttel2017ActaNumerica, Martin-Reining-Ceperley2016book}, where the self-energy has been
assumed to be diagonal in the reference basis. This reduces the problem to a scalar but still nonlinear equation for each state, possibly leading to multiple solutions.
To simplify the problem, one can linearize the self-energy as the first-order term of the Taylor expansion around $\varepsilon_{n\kk}$, yielding:
\begin{equation}
\varepsilon_{n\kk}^{\mathrm{QP}} =
\varepsilon_{n\kk} +
Z_{n\kk}\,
\mel{n\kk}{\Sigma(\varepsilon_{n\kk}) - v_{xc}}{n\kk},
\label{eq_QP_lin}
\end{equation}
with the renormalization factor
\begin{equation}
Z_{n\kk} =
\left[
1 -
\left.
\mel{n\kk}{\frac{\partial\Sigma(\omega)}{\partial\omega}}{n\kk}
\right|_{\omega=\varepsilon_{n\kk}}
\right]^{-1}.
\label{Eq_QP_Z}
\end{equation}

Equations~\eqref{Eq_QP_full} or \eqref{eq_QP_lin} can be solved numerically by using iterative root-finding methods. 
Multiple solutions have been reported, for instance, for some molecules in the GW100 benchmark~\cite{gw100}. 
When both the real and imaginary parts of $\Sigma$ are considered, the different solutions correspond to the poles of the interacting Green’s function $G$, and can be assigned one to the main QP excitation and the others to satellites. 
The solutions can also be determined analytically using alternative approaches based on rational representations of $\Sigma$,  through, for example, the algorithmic inversion method for sum-over-poles (AIM-SOP) of Refs.~\cite{Chiarotti2022PRR,Chiarotti2024PRR,Ferretti2024PRB,Quinzi2025PRB}, or the MPA-$\Sigma$ approach of Ref.~\cite{Leon2025PhysRevB}.

In this work, we adopt the linearized form of Eq.~\eqref{eq_QP_lin}, solved using Newton’s method as implemented in the \yambo{} code.  
This approach assumes that $\Sigma$ varies smoothly around the QP energy and therefore neglects secondary (satellite) solutions.

\subsection{GW self-energy evaluation}
%

The state-of-the-art method for evaluating the self-energy $\Sigma$ is the one-shot $G_0W_0$ approximation. In this approximation, $\Sigma$ is given by a frequency convolution of the non-interacting KS Green's function $G_0$ and the screened Coulomb potential $W_0$:
\begin{equation}
    \Sigma(\omega) = \frac{i}{2 \pi} \int{d\omega' e^{i \omega' 0^+} G_0(\omega+\omega')W_0(\omega')}.
    \label{eq_sigma_convolution}
\end{equation}
In the expression above, $W_0$ is usually evaluated at the random-phase approximation (RPA) level and is typically split into two terms: the bare Coulomb interaction $v$, which is static, and the correlation part $W^c\equiv W_0-v$, which contains all the frequency dependence. This, in turn, leads to a decomposition of the self-energy into exchange and correlation terms, $\Sigma(\o) = \Sigma^x + \Sigma^c(\o)$.

The dynamical nature of the screened interaction $W$ distinguishes the GW approximation from static theories like Hartree-Fock.
A number of different strategies have been proposed to evaluate the frequency integral in Eq.~\eqref{eq_sigma_convolution}, ranging from full-analytic~\cite{van_setten_gw_2013} (FA) to full-frequency real-axis integration~\cite{liu_numerical_2015} (FF), analytic continuation~\cite{rieger_gw_1999} (AC), contour deformation (CD)~\cite{lebegue_implementation_2003}, and multipole approaches (MPA)~\cite{Valido_2021,leon2023efficient}. The earliest and simplest approach is the so called plasmon-pole approximation~\cite{hybertsen_electron_1986,godby_metal-insulator_1989} (PPA).
Within PPA, the frequency dependence of $W^c$ is simplified as a symmetric single pole function for each matrix element $\G\G'$ and transferred momentum $\q$:
\begin{equation}
    W^{c\mathrm{PPA}}_{\G\G'}(\q,\omega) = \frac{2 R_{\q\G\G'} \Omega_{\q\G\G'}}{\omega^2-\Omega_{\q\G\G'}^2} .
    \label{eq:W_PPA}
\end{equation}

Among the different flavors of PPA~\cite{stankovski_g_2011}, in this work we  adopt the one proposed by Godby and Needs (GN-PPA)~\cite{godby_metal-insulator_1989}, which interpolates $W^c$ from the values computed in two points, $\omega = 0$ and $\omega = i E_{\mathrm{PPA}}$. 
Here, we set E$_{\text{PPA}}$= 30 eV. 
The Hybertsen-Louie generalized plasmon-pole model~\cite{hybertsen_electron_1986} (HL-PPA)
follows a different approach and, in addition to evaluating the screening at zero frequency, imposes
the fulfillment of Johnson's frequency sum rule (\textit{f}-sum rule)~\cite{f_sum}. 
Although such a property can be mapped with the GN-PPA model, by increasing $E_{\mathrm{PPA}}$ to infinity~\cite{Valido_2021}, the results of the standard GN-PPA and HL-PPA may differ significantly. For a detailed comparison of different PPA models, see, e.g., Refs.~[\citenum{stankovski_g_2011,Miglio2012EPJB,Larson2013PRB}].

Besides GN-PPA, we also adopt the multipole approximation (MPA) method~\cite{Valido_2021,leon2023efficient}, which generalizes Eq.~\eqref{eq:W_PPA} to multiple complex poles:
\begin{equation}
    W^{c\mathrm{MPA}}_{\G\G'} (\q, \omega) = \sum_p^{n_p} \frac{2 R_{p\q\G\G'} \Omega_{p\q\G\G'}}{\omega^2-\Omega_{p\q\G\G'}^2},
    \label{eq:W_MPA}
\end{equation}
where $n_p$ is typically around 10.
Making use of the Lehmann representation for $G_0$ in a Bloch plane-wave basis set, and the MPA model for $W^c$, the diagonal matrix elements of the correlation self-energy are integrated analytically and can be expressed as
\begin{multline}
    \Sigma^{c\mathrm{MPA}}_{n \kk} (\omega) =\sum_m \sum_{\G \G'} \sum_p^{n_p} \int \frac{d\q}{(2\pi)^3} S^{n m}_{p \G \G'} (\kk,\q) \times
    \\
   \left [ \frac{f_{m\kk-\q}^{\text{KS}}}
     {\o-\varepsilon_{m\kk-\q}^{\text{KS}}+\Omega_{p\q\G\G'}} + \frac{1-f_{m\kk-\q}^{\text{KS}}}
     {\o-\varepsilon_{m\kk-\q}^{\text{KS}}-\Omega_{p\q\G\G'}} \right ],
    \label{eq:Sc}
\end{multline}
where we have defined
\begin{eqnarray}
    S^{nm}_{p\G\G'}(\kk,\q) &=& -2 \rho_{nm}^{\text{KS}}(\kk,\q,\G)R_{p\q\G\G'}{\rho_{nm}^{\text{KS}}}^*(\kk,\q,\G'), 
    \nonumber
    \\
    \rho_{nm}^{\text{KS}}(\kk,\q,\G)  &=& \mel{n\mathbf{k}}{e^{i(\mathbf{q}+\mathbf{G}) \cdot \mathbf{r}}}{m\mathbf{k-q}}.
\end{eqnarray}

For each momentum transfer $\bf q$ and each $\bf GG'$ matrix element, poles and residues are obtained through a non-linear interpolation with frequency points $z_i$ 
sampled in the complex plane:
\begin{equation}
    \sum\limits_{p=1}^{n_p}
    \frac{2 R_{p\q\G\G'}\O_{p\q\G\G'}}{z_i^2-\O_{p\q\G\G'}^2} = W^{\text{c}}_ {\G\G'}(\q,z_i), \ 
    \label{Eq_MPA_av}
\end{equation}
where $i=1,\ldots,2n_p$ and the set of complex frequencies $z_i$ are conveniently selected according to the double parallel sampling defined in Refs.~\cite{Valido_2021,leon2023efficient}.
Such efficient sampling allows the MPA method to achieve FF accuracy with a small number of poles or sampling frequencies, typically one to two orders of magnitude fewer than the number of frequency points required to converge FF-RA.

\subsection{Convergences and workflow}
\label{sec:workflow}

\begin{figure}
    \centering
    \includegraphics[width=0.95\columnwidth]{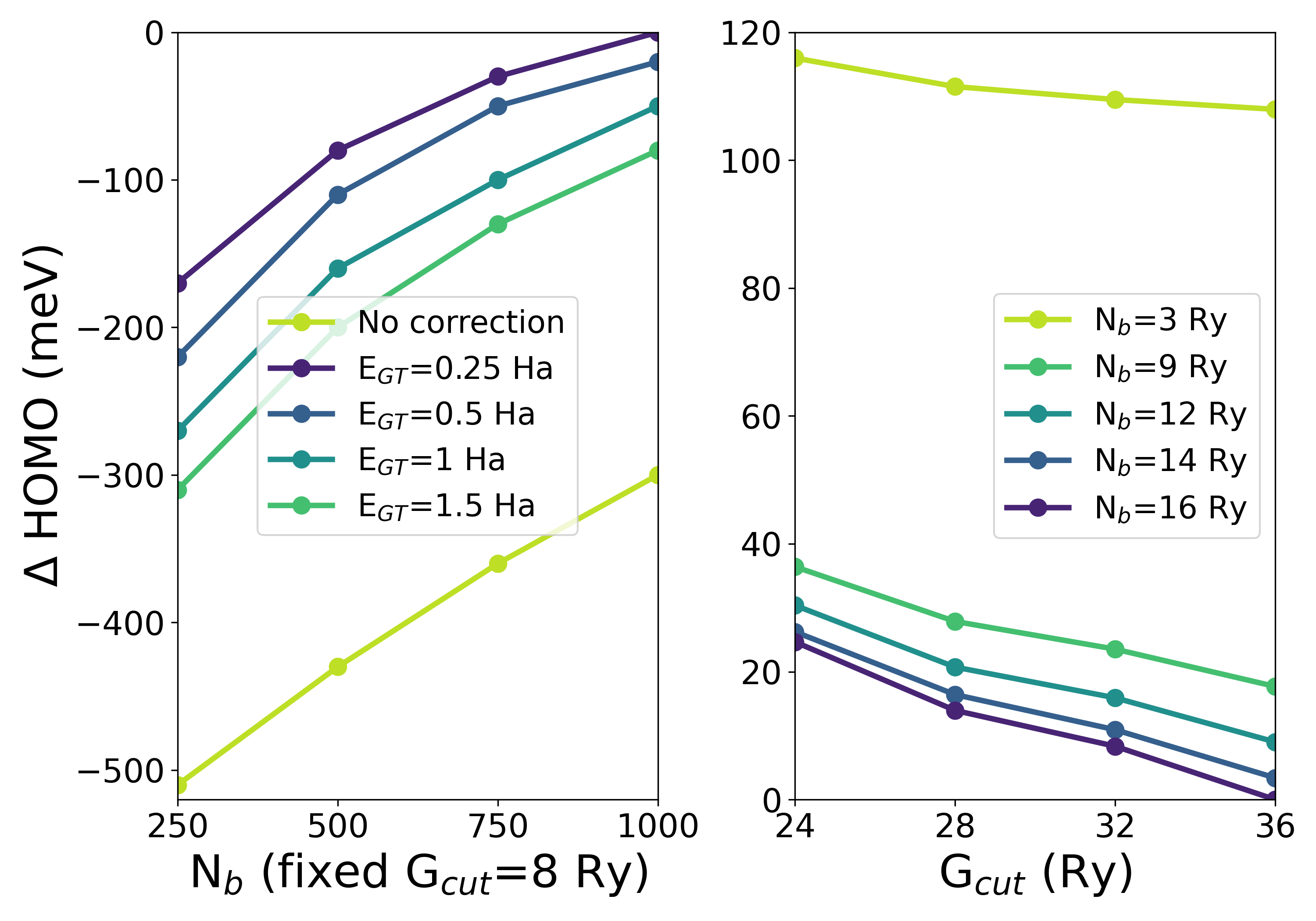}
    \caption{Convergence tests for the LiF molecule. Left panel: Effect of E$_{GT}$ (see text) on the convergence acceleration of the HOMO with respect to empty states summation (the value at N$_b$=1000 for E$_{GT}$=0.25 Ha is set to zero in the plot). Right panel: HOMO energy versus the PW cutoff $G_{cut}$ in the screening matrix, for different number of empty states, N$_b$. The result obtained with N$_b$=16 Ry $G_{cut}$=36 Ry is set to zero. The values N$_b$ = 3, 9, 12, 14, 16 Ry correspond to 1000, 5000, 7000, 9000 and 11000 bands.}
    \label{fig:Lif}
\end{figure}

Accurate GW QP energies require convergence with respect to several computational parameters. In \yambo{}, the most relevant parameters are the number of unoccupied states ($N_b$) that enter the sum over states and the kinetic-energy cutoff ($G_{cut}$) defining the number of reciprocal lattice vectors used in the evaluation of the self-energy in Eq.~\eqref{eq:Sc} and the screening matrix in Eq.~\eqref{eq:W_MPA}.
Within PPA, convergence with respect to $N_b$ can be accelerated by using the Bruneval–Gonze (BG) terminator~\cite{BG}, which replaces the contribution of high-energy states with an effective pole at an energy $E_{GT}$ above the highest explicitly included band. For isolated systems, where unoccupied levels are densely spaced, effective $E_{GT}$ values are typically smaller than for bulk systems. Tests on LiF (Fig.~\ref{fig:Lif}, left panel) show that $E_{GT}=0.25$~Ha yields optimal convergence, and this value is adopted for all molecules. 

As discussed in the literature~\cite{bonacci2023,rangel_repr_2020,stankovski_g_2011}, the parameters $N_b$ and $G_{cut}$ are interdependent, as shown in the right panel of Fig.~\ref{fig:Lif}, where the curves indicate that the quasiparticle energies have not yet reached convergence within the accessible range of parameters. In this work, the convergence with respect to these two parameters is assessed simultaneously via extrapolation. Specifically, QP corrections are computed on a two-dimensional grid with $N_b$ ranging from 1000 to 11000 (3000-13000 for MPA) and $G_{cut}$ from 24 to 36~Ry. When possible, an additional point at $G_{cut}=38$~Ry is included.
The extrapolation to infinite cutoff values is obtained by fitting the QP corrections to:
\begin{equation}
    f(N_b,G_{cut}) = \left( \frac{A}{N_b^{\alpha}} + b \right)
                     \left( \frac{C}{G_{cut}^{\beta}} + d \right),
    \label{multi_over_x_}
\end{equation}
where $\alpha$ and $\beta$ belong to $\{1,2,3\}$ and, for each molecule, the values that minimize the root-mean-square error are chosen. 
Similarly to the strategy adopted in refs.~\cite{schindlmayr_analytic_2013,klimes_predictive_2014,rasmussen_computational_2015,van_setten_automation_2017,rasmussen_towards_2021}, the fully converged QP energy reads:

\begin{equation}
    E^{\mathrm{QP}}_{\mathrm{extra}} = f(N_b \to \infty, G_{cut} \to \infty) = b d.
    \label{extra}
\end{equation}

All calculations were performed with a fully automated high-throughput (HT) scheme using the \texttt{aiida-yambo} plugin~\cite{bonacci2023,yambo-aiida-github}, which is part of the AiiDA workflow ecosystem~\cite{huber_aiida_2020,uhrin_workflows_2021} and implements automated workflows and convergence protocols for MBPT \yambo{} simulations. We note that related efforts have been undertaken by the community in recent years~\cite{biswas_pygwbse:_2023,grosmann_robust_2024,varrassi_automated_2025}. For each of the 100 molecules, convergence workflows systematically explored the $(N_b,G_{cut})$ space, resulting in over 9000 QP evaluations across both PPA and MPA schemes. All input and output data are stored in the AiiDA database, ensuring full reproducibility. This work also serves as a large-scale validation of the \texttt{aiida-yambo} workflows. Computational performance statistics are reported in the \suppinfo.

\section{Results}
\label{sec:results}
%
To benchmark the accuracy of the GN–PPA and MPA implementations of the $G_0W_0$ method in \yambo{}, as compared to other codes, we computed the IP and EA energies of the 100 molecules in the GW100 dataset.  
These quantities are evaluated as
\begin{equation}
\begin{split}
    \mathrm{IP} &= E^{\mathrm{vac}} - E^{\mathrm{HOMO}}, \\
    \mathrm{EA} &= E^{\mathrm{vac}} - E^{\mathrm{LUMO}},
\end{split}
\end{equation}
where $E^{\mathrm{vac}}$ denotes the electrostatic potential in the vacuum region, evaluated at the DFT level.

The statistical accuracy of the computed IP energies is characterized by the mean error (ME), the mean absolute error (MAE), the mean absolute relative error (MARE), and the standard deviation ($\sigma$), summarized in Table~\ref{tab:MAE} of Sec.~\ref{sec:IP}. The  IP values obtained for each molecule are reported in Table~\ref{tab:HOMO_RES_1}, together with reference GW results~\cite{gw100} from other codes, CCSD data~\cite{CCSD}, and experimental values from the NIST database~\cite{NIST_db}. EA results are provided in Table~\ref{tab:LUMO_RES} and discussed in Sec.~\ref{sec:EA}.
While the discussion below focuses primarily on IPs, which provide the most direct and comprehensive benchmark across the different implementations, the same conclusions apply to the EAs, analyzed in Sec.~\ref{sec:EA}.

\subsection{Ionization Potentials}
\label{sec:IP}
%

%
\begin{table*}[]
\begin{tabularx}{\textwidth}{
l|
>{\centering\arraybackslash}X
>{\centering\arraybackslash}X|
>{\centering\arraybackslash}X
>{\centering\arraybackslash}X|
>{\centering\arraybackslash}X
>{\centering\arraybackslash}X|
>{\centering\arraybackslash}X
>{\centering\arraybackslash}X}
\hline\hline
    &&&&&&&&\\[-7pt]
\textbf{Code} & \multicolumn{2}{c|}{\textbf{ME} (meV)}   & \multicolumn{2}{c|}{\textbf{MAE} (meV)}   & \multicolumn{2}{c|}{\textbf{MARE} ($\%$)} & \multicolumn{2}{c}{ $\boldsymbol{\sigma}$ (meV)}\\[4pt]
    & GN-PPA & MPA &GN-PPA & MPA &GN-PPA & MPA &GN-PPA & MPA \\[4pt]
    \hline
    &&&&&&&&\\[-7pt]
    YAMBO$_\text{GN-PPA}$ & - & 77 & - & 168 & - & 1.8 & - & 114 \\[4pt]
    WEST$^\mathrm{extra}_\mathrm{lin}$ & -117 & -31 & 176 & 108 & 1.8 & 1.3 & 159 & 121 \\[4pt]
    WEST$^\mathrm{extra}_\mathrm{sol}$ & 9 & 96 & 159 & 167 & 1.8 & 1.9 & 165 & 123 \\[4pt]
    VASP & -44 & 32 & 234 & 155 & 2.4 & 1.8 & 216 & 199 \\[4pt]
    AIMS$^\mathrm{pade}$ & 225 & 302 & 299 & 334 & 3.4 & 3.6 & 320 & 260 \\[4pt]
    AIMS$^\mathrm{extra}$ & 22 & 119 & 214 & 189 & 2.4 & 2.1 & 200 & 185 \\[4pt]
    BGW$_\text{HL-PPA}$ & -400 & -323 & 511 & 400 & 4.7 & 3.8 & 336 & 321 \\[4pt]
    BGW$_\text{FF}$ & 203 & 387 & 262 & 399 & 2.8 & 3.9 & 382 & 388 \\[4pt]
    CCSD & -338 & -259 & 447 & 350 & 4.1 & 3.3 & 374 & 310 \\[4pt]
    EXP & -255 & -175 & 534 & 454 & 5.0 & 4.3 & 401 & 387 \\[4pt]
\hline\hline
\end{tabularx}
\caption{Mean error (ME), mean absolute error (MAE), mean absolute relative error (MARE) and standard deviation ($\sigma$) of the IPs computed with \yambo-\text{GN-PPA} and \yambo-{MPA} with respect to reference GW100 results~\cite{gw100, maggio+gwsmall, gov+west+gw100} obtained with different codes and approximations. We also included experimental (EXP) data~\cite{gw100,NIST_db}.
}
\label{tab:MAE}
\end{table*}
\begin{table*}[]
\centering
\begin{adjustbox}{max width=\linewidth}
\begin{tabular}{cl|ccccccccccc}
\hline\hline
    &\\[-5pt]
index & Formula & Y$_\text{MPA}$ & Y$_\text{PPA}$ &
WEST$_\mathrm{lin}^{extra}$ & WEST$_\mathrm{sol}^\mathrm{extra}$ & VASP & AIMS$^\mathrm{pade}$ & AIMS$^\mathrm{extra}$ & BGW$_\text{HL}$ & BGW$_\text{FF}$ & CCSD & EXP\\
& \\[-7pt]
\hline\hline\\[-9pt]
1 & He & 23.72 & 23.33 & 23.65 & 23.42 & 23.62 & 23.48 & 23.49 & 24.1 & - & 24.51 & 24.59 \\
2 & Ne & 20.14 & 20.06 & 20.52 & 20.33 & 20.36 & 20.38 & 20.33 & 21.35 & - & 21.32 & 21.56 \\
3 & Ar & 15.56 & 15.34 & 15.5 & 15.37 & 15.42 & 15.13 & 15.28 & 15.94 & - & 15.54 & 15.76 \\
4 & Kr & 13.9 & 13.79 & 13.87 & 13.76 & 14.03 & 13.57 & 13.89 & 14.0 & - & 13.94 & 14.0 \\
5 & Xe & 13.34 & 13.52 & 13.38 & 13.22 & 12.22 & 12.02 & - & 12.08 & - & - & 12.13 \\
6 & H$_2$ & 16.1 & 16.05 & 16.03 & 15.84 & 16.06 & 15.82 & 15.85 & 16.23 & - & 16.4 & 15.43 \\
7 & Li$_2$ & 5.07 & 5.24 & 5.19 & 5.04 & 5.32 & 4.99 & 5.05 & 5.43 & - & 5.27 & 4.73 \\
8 & Na$_2$ & 4.89 & 5.11 & 5.07 & 4.98 & 5.06 & 4.83 & 4.88 & 5.03 & - & 4.95 & 4.89 \\
9 & Na$_4$ & 3.85 & 4.09 & 4.28 & 4.24 & 4.23 & 4.1 & 4.14 & 4.34 & - & 4.22 & 4.27 \\
10 & Na$_6$ & 3.78 & 4.04 & 4.42 & 4.37 & 4.4 & 4.24 & 4.34 & 4.47 & - & 4.35 & 4.12 \\
11 & K$_2$ & 3.96 & 4.3 & 4.21 & 4.14 & 4.24 & 3.98 & 4.08 & 4.02 & - & 4.06 & 4.06 \\
12 & Rb$_2$ & 3.79 & 4.14 & 4.08 & 4.01 & 4.14 & 3.8 & - & 3.92 & - & 3.92 & 3.9 \\
13 & N$_2$ & 15.16 & 14.84 & 15.08 & 14.94 & 15.06 & 14.89 & 15.05 & 15.43 & 14.72 & 15.57 & 15.58 \\
14 & P$_2$ & 10.53 & 10.57 & 10.48 & 10.43 & 10.4 & 10.21 & 10.38 & 10.66 & - & 10.47 & 10.62 \\
15 & As$_2$ & 9.62 & 9.7 & 9.58 & 9.55 & 9.62 & 9.47 & 9.67 & 9.67 & - & 9.78 & 10.0 \\
16 & F$_2$ & 14.89 & 14.53 & 15.16 & 15.0 & 15.08 & 14.96 & 15.1 & 15.59 & 14.73 & 15.71 & 15.7 \\
17 & Cl$_2$ & 11.61 & 11.45 & 11.5 & 11.41 & 11.4 & 11.1 & 11.31 & 11.85 & - & 11.41 & 11.49 \\
18 & Br$_2$ & 10.62 & 10.5 & 10.52 & 10.44 & 10.65 & 10.22 & 10.56 & 10.64 & - & 10.54 & 10.51 \\
19 & I$_2$ & 10.41 & 10.61 & 10.56 & 10.41 & 9.59 & 9.28 & - & 9.58 & - & 9.51 & 9.36 \\
20 & CH$_4$ & 14.2 & 14.03 & 14.1 & 13.99 & 14.14 & 13.93 & 14.0 & 14.28 & 13.8 & 14.37 & 13.6 \\
21 & C$_2$H$_6$ & 12.63 & 12.44 & 12.53 & 12.44 & 12.58 & 12.36 & 12.46 & 12.63 & 12.22 & 13.04 & 11.99 \\
22 & C$_3$H$_8$ & 12.04 & 11.85 & 11.92 & 11.84 & 11.98 & 11.79 & 11.89 & 12.05 & - & 12.05 & 11.51 \\
23 & C$_4$H$_10$ & 11.7 & 11.52 & 11.48 & 11.41 & 11.69 & 11.49 & 11.59 & 11.73 & - & 11.57 & 11.09 \\
24 & C$_2$H$_4$ & 10.52 & 10.48 & 10.46 & 10.39 & 10.5 & 10.32 & 10.4 & 10.68 & 10.3 & 10.67 & 10.68 \\
25 & C$_2$H$_2$ & 11.28 & 11.18 & 11.18 & 11.09 & 11.24 & 11.02 & 11.09 & 11.35 & 10.97 & 11.42 & 11.49 \\
26 & C$_4$ & 11.11 & 10.98 & 10.97 & 10.9 & 10.97 & 10.78 & 10.91 & 11.49 & - & 11.26 & 12.54 \\
27 & C$_3$H$_6$ & 10.78 & 10.63 & 10.73 & 10.67 & 10.78 & 10.56 & 10.65 & 10.93 & - & 10.86 & 10.54 \\
28 & C$_6$H$_6$ & 9.17 & 9.1 & 9.13 & 9.08 & 9.16 & 8.99 & 9.1 & 9.21 & - & 9.29 & 9.23 \\
29 & C$_8$H$_8$ & 8.21 & 8.09 & 8.2 & 8.16 & 8.24 & 8.06 & 8.18 & 8.47 & - & 8.35 & 8.43 \\
30 & C$_5$H$_6$ & 8.56 & 8.48 & 8.49 & 8.44 & 8.51 & 8.35 & 8.45 & 8.77 & - & 8.68 & 8.53 \\
31 & C$_2$H$_3$F & 10.4 & 10.28 & 10.36 & 10.29 & 10.36 & 10.2 & 10.32 & 10.8 & 10.14 & 10.55 & 10.63 \\
32 & C$_2$H$_3$Cl & 10.09 & 9.97 & 10.0 & 9.94 & 10.0 & 9.76 & 9.9 & 10.32 & - & 10.09 & 10.2 \\
33 & C$_2$H$_3$Br & 9.23 & 9.17 & 9.71 & 9.64 & 9.83 & 8.99 & 9.14 & 9.42 & - & 9.27 & 9.9 \\
34 & C$_2$H$_3$I & 9.85 & 9.9 & 9.94 & 9.81 & 9.36 & 9.04 & - & 9.48 & - & 9.33 & 9.35 \\
35 & CF$_4$ & 15.46 & 15.13 & 15.65 & 15.51 & 15.53 & 15.36 & 15.6 & 15.96 & - & 16.3 & 16.2 \\
36 & CCl$_4$ & 11.49 & 11.35 & 11.41 & 11.29 & 11.31 & 10.98 & 11.21 & 11.77 & - & 11.56 & 11.69 \\
37 & CBr$_4$ & 10.25 & 10.15 & 10.22 & 10.11 & 10.38 & 9.9 & 10.22 & 10.4 & - & 10.46 & 10.54 \\
38 & CI$_4$ & 9.82 & 10.02 & - & - & 9.23 & 8.82 & - & 9.23 & - & 9.27 & 9.1 \\
39 & SiH$_4$ & 12.6 & 12.6 & 12.55 & 12.42 & 12.53 & 12.31 & 12.4 & 12.77 & - & 12.8 & 12.3 \\
40 & GeH$_4$ & 12.47 & 12.53 & 12.44 & 12.32 & 12.24 & 12.02 & 12.12 & 12.28 & - & 12.5 & 11.34 \\
41 & Si$_2$H$_6$ & 10.6 & 10.6 & 10.58 & 10.52 & 10.52 & 10.31 & 10.41 & 10.8 & - & 10.64 & 10.53 \\
42 & Si$_5$H$_12$ & 8.69 & 8.7 & 9.25 & 9.19 & 9.19 & 8.94 & 9.05 & 9.45 & - & 9.27 & 9.36 \\
43 & LiH & 7.31 & 7.55 & 7.2 & 6.62 & 7.2 & 6.54 & 6.58 & 7.85 & 6.67 & 7.96 & 7.9 \\
44 & KH & 5.55 & 5.99 & 5.39 & 4.97 & 5.37 & 4.86 & 4.99 & 5.76 & - & 6.13 & 8.0 \\
45 & BH$_3$ & 13.16 & 13.07 & 13.08 & 12.95 & 13.09 & 12.87 & 12.96 & 13.28 & - & 13.28 & 12.03 \\
46 & B$_2$H$_6$ & 12.1 & 12.0 & 12.03 & 11.92 & 12.04 & 11.84 & 11.93 & 12.17 & - & 12.26 & 11.9 \\
47 & NH$_3$ & 10.46 & 10.3 & 10.4 & 10.18 & 10.44 & 10.32 & 10.39 & 10.93 & - & 10.81 & 10.82 \\
48 & HN$_3$ & 10.63 & 10.35 & 10.54 & 10.48 & 10.56 & 10.4 & 10.55 & 10.96 & - & 10.68 & 10.72 \\
49 & PH$_3$ & 10.57 & 10.61 & 10.51 & 10.43 & 10.45 & 10.27 & 10.36 & 10.79 & - & 10.52 & 10.59 \\
50 & AsH$_3$ & 10.43 & 10.45 & 10.4 & 10.33 & 10.36 & 10.12 & 10.21 & 10.45 & - & 10.4 & 10.58 \\ \\
    \hline\hline
\end{tabular}
\end{adjustbox}
\caption{Quasiparticle ionization potentials (IP) of the molecules in the GW100 dataset as obtained within this work: Y$_\text{PPA}$ and  Y$_\text{MPA}$ refer to \yambo{} calculations performed at the PPA and MPA level, respectively; For completeness, PBE IPs are also reported. Reference GW results from other codes are also provided (AIMS, TM, BGW from Ref.~\cite{gw100}, and WEST from Ref.~\cite{gw100}), together with experimental data~\cite{NIST_db}.
}
\label{tab:HOMO_RES_1}
\end{table*}
\addtocounter{table}{-1}
\begin{table*}[]
\centering
\begin{adjustbox}{max width=\linewidth}
\begin{tabular}{cl|ccccccccccc}
\hline\hline
    &\\[-5pt]
index & Formula & Y$_\text{MPA}$ & Y$_\text{PPA}$ &
WEST$_\mathrm{lin}^\mathrm{extra}$ & WEST$_\mathrm{sol}^\mathrm{extra}$ & VASP & AIMS$^\mathrm{pade}$ & AIMS$^\mathrm{extra}$ & BGW$_\text{HL}$ & BGW$_\text{FF}$ & CCSD & EXP\\
& \\[-7pt]
\hline\hline\\[-9pt]

51 & SH$_2$ & 10.43 & 10.37 & 10.36 & 10.23 & 10.3 & 10.03 & 10.13 & 10.64 & - & 10.31 & 10.5 \\
52 & FH & 15.34 & 15.03 & 15.47 & 15.23 & 15.38 & 15.3 & 15.37 & 16.24 & - & 16.03 & 16.12 \\
53 & ClH & 12.68 & 12.58 & 12.6 & 12.48 & 12.51 & 12.25 & 12.36 & 12.97 & - & 12.59 & 12.79 \\
54 & LiF & 10.36 & 10.32 & 10.54 & 10.11 & 10.45 & 9.95 & 10.27 & 11.84 & - & 11.32 & 11.3 \\
55 & F$_2$Mg & 12.65 & 12.59 & 12.84 & 12.46 & 12.77 & 12.32 & 12.5 & 13.73 & 12.44 & 13.71 & 13.3 \\
56 & TiF$_4$ & 14.12 & 13.95 & 14.31 & - & 14.22 & 13.89 & 14.07 & 14.88 & - & 15.48 & 13.3 \\
57 & AlF$_3$ & 14.45 & 14.22 & 14.63 & 14.4 & 14.53 & 14.25 & 14.48 & 15.11 & - & 15.46 & 15.45 \\
58 & BF & 10.69 & 11.08 & 10.71 & 10.56 & 10.67 & 10.56 & 10.73 & 11.49 & - & 11.09 & 11.0 \\
59 & SF$_4$ & 12.43 & 12.4 & 12.41 & 12.32 & 12.29 & 12.12 & 12.38 & 12.79 & - & 12.59 & 11.69 \\
60 & BrK & 8.04 & 8.1 & 7.96 & - & 8.04 & 7.3 & 7.57 & 7.99 & - & 8.13 & 8.82 \\
61 & GaCl & 10.35 & 10.29 & 10.27 & 10.19 & 9.99 & 9.55 & 9.74 & 10.24 & - & 9.77 & 10.07 \\
62 & NaCl & 8.98 & 9.02 & - & - & 8.76 & 8.1 & 8.43 & 9.6 & - & 9.03 & 9.8 \\
63 & MgCl$_2$ & 11.46 & 11.45 & 11.47 & 11.25 & 11.41 & 10.99 & 11.2 & 11.98 & - & 11.66 & 11.8 \\
64 & AlI$_3$ & 10.12 & 10.3 & 10.59 & 10.31 & 9.69 & 9.32 & - & 9.67 & - & 9.82 & 9.66 \\
65 & BN & 11.6 & 11.34 & - & - & 10.61 & 11.03 & 11.15 & 12.19 & 9.68 & 11.89 & 11.5 \\
66 & NCH & 13.5 & 13.3 & 13.41 & 13.22 & 13.43 & 13.21 & 13.32 & 13.87 & - & 13.87 & 13.61 \\
67 & PN & 11.51 & 11.28 & 11.43 & 11.26 & 11.41 & 11.14 & 11.29 & 12.13 & - & 11.74 & 11.88 \\
68 & H$_2$NNH$_2$ & 9.48 & 9.21 & 9.42 & 9.27 & 9.45 & 9.28 & 9.37 & 9.78 & 9.1 & 9.72 & 8.98 \\
69 & H$_2$CO & 10.57 & 10.24 & 10.56 & 10.41 & 10.57 & 10.33 & 10.46 & 11.02 & - & 10.84 & 10.88 \\
70 & CH$_4$O & 10.64 & 10.39 & 10.74 & 10.6 & 10.72 & 10.56 & 10.67 & 11.14 & - & 11.04 & 10.96 \\
71 & C$_2$H$_6$O & 10.36 & 10.0 & 10.36 & 10.21 & 10.33 & 10.16 & 10.27 & 10.57 & - & 10.68 & 10.64 \\
72 & C$_2$H$_4$O & 9.84 & 9.46 & 9.81 & 9.61 & 9.8 & 9.55 & 9.66 & 10.16 & 9.43 & 10.21 & 10.24 \\
73 & C$_4$H$_10$O & 9.46 & 9.13 & 9.52 & 9.4 & 9.52 & 9.32 & 9.42 & 9.7 & - & 9.82 & 9.61 \\
74 & CH$_2$O$_2$ & 10.87 & 10.72 & 11.01 & 10.82 & 10.98 & 10.73 & 10.87 & 11.39 & - & 11.42 & 11.5 \\
75 & HOOH & 11.18 & 10.79 & 11.16 & 11.0 & 11.12 & 10.99 & 11.1 & 11.58 & 10.82 & 11.59 & 11.7 \\
76 & H$_2$O & 12.08 & 11.81 & 12.09 & 11.87 & 12.05 & 11.97 & 12.05 & 12.75 & 11.68 & 12.56 & 12.62 \\
77 & CO$_2$ & 13.45 & 13.1 & 13.46 & 13.37 & 13.44 & 13.25 & 13.46 & 13.81 & 13.17 & 13.71 & 13.77 \\
78 & CS$_2$ & 10.15 & 10.09 & 10.1 & 10.05 & 10.01 & 9.75 & 9.95 & 10.37 & - & 9.98 & 10.09 \\
79 & OCS & 11.26 & 11.15 & - & - & 11.13 & 10.91 & 11.11 & 11.49 & 11.02 & 11.17 & 11.19 \\
80 & OCSe & 10.49 & 10.45 & 10.43 & 10.37 & 10.5 & 10.2 & 10.43 & 10.55 & - & 10.78 & 10.37 \\
81 & CO & 13.81 & 13.77 & 13.79 & 13.66 & 13.76 & 13.57 & 13.71 & 14.33 & - & 14.21 & 14.01 \\
82 & O$_3$ & 12.18 & 11.84 & - & - & 12.07 & 11.39 & 11.49 & 13.05 & 12.0 & 12.55 & 12.73 \\
83 & SO$_2$ & 12.08 & 11.81 & 12.08 & 11.96 & 12.04 & 11.82 & 12.06 & 12.55 & - & 13.49 & 12.5 \\
84 & BeO & 9.57 & 9.57 & - & - & 9.5 & 8.58 & 8.6 & 10.66 & 9.68 & 9.94 & 10.1 \\
85 & MgO & 7.34 & 7.32 & - & - & 7.1 & 6.68 & 6.75 & 8.51 & 7.08 & 7.49 & 8.76 \\
86 & C$_7$H$_8$ & 8.69 & 8.65 & 8.75 & 8.71 & 8.79 & 8.61 & 8.72 & 8.97 & - & 8.9 & 8.82 \\
87 & C$_8$H$_10$ & 8.61 & 8.52 & 8.7 & 8.66 & 8.73 & 8.55 & 8.66 & 8.92 & - & 8.85 & 8.77 \\
88 & C$_6$F$_6$ & 9.61 & 9.46 & 9.7 & 9.65 & 9.69 & 9.49 & 9.74 & 10.04 & - & 9.93 & 10.2 \\
89 & C$_6$H$_5$OH & 8.4 & 8.36 & 8.42 & 8.37 & 8.43 & 8.37 & 8.51 & 8.72 & - & 8.7 & 8.75 \\
90 & C$_6$H$_5$NH$_2$ & 7.79 & 7.64 & 7.8 & 7.73 & 7.84 & 7.64 & 7.78 & 7.98 & - & 7.99 & 8.05 \\
91 & C$_5$H$_5$N & 9.32 & 9.1 & 9.28 & 9.13 & 9.31 & 9.04 & 9.17 & 9.5 & - & 9.66 & 9.66 \\
92 & C$_5$H$_5$N$_5$O & 7.71 & 7.54 & 7.86 & 7.82 & 8.18 & 7.69 & 7.87 & 7.92 & - & 8.03 & 8.24 \\
93 & C$_5$H$_5$N$_5$ & 8.08 & 7.9 & 8.14 & 8.09 & 8.18 & 7.98 & 8.15 & 8.35 & - & 8.33 & 8.48 \\
94 & C$_4$H$_5$N$_3$O & 8.46 & 8.26 & 8.49 & 8.4 & 8.5 & 8.29 & 8.44 & 8.77 & - & 9.51 & 8.94 \\
95 & C$_5$H$_6$N$_2$O$_2$ & 8.81 & 8.62 & 8.88 & 8.82 & 8.89 & 8.71 & 8.87 & 9.19 & - & 9.08 & 9.2 \\
96 & C$_4$H$_4$N$_2$O$_2$ & 9.38 & 9.22 & 9.26 & 9.19 & 9.55 & 9.22 & 9.38 & 9.94 & - & 10.12 & 9.68 \\
97 & CH$_4$N$_2$O & 9.58 & 9.33 & 9.6 & 9.4 & 9.59 & 9.32 & 9.46 & 9.94 & - & 10.05 & 9.8 \\
98 & Ag$_2$ & 8.11 & 8.27 & 8.12 & 8.04 & 7.95 & 7.07 & - & 8.57 & - & 7.49 & 7.66 \\
99 & Cu$_2$ & 7.24 & 7.66 & - & - & 7.4 & 7.54 & 7.78 & 8.6 & - & 7.57 & 7.46 \\
100 & NCCu & 9.93 & 10.12 & - & - & 9.99 & 9.42 & 9.56 & 10.91 & - & 10.85 & - \\ \\
    \hline\hline
\end{tabular}
\end{adjustbox}
\caption{Continuation of Table \ref{tab:HOMO_RES_1}.}
\label{tab:HOMO_RES_2}
\end{table*}
%
%

We first examined the computed Kohn–Sham (PBE) HOMO energies, which serve as starting point for the subsequent $GW$ quasiparticle corrections. These results are reported in Sec.~II of the \suppinfo. The agreement with previous \textsc{Quantum ESPRESSO} calculations \cite{gov+west+gw100} is very good; the remaining small discrepancies arise primarily from the different supercell shape adopted in this work (see Sec.~\ref{sec:setup}).
Compared with results obtained using localized basis set codes, such as \textsc{FHI-aims} or \textsc{TM}, the differences reflect both the different basis representations and the use of pseudopotentials, which influence the high-energy part of the spectrum.
Interestingly, the present PBE HOMO values agree well (within 40 meV on average) with def2-QZVP GTO results (see Ref.~\cite{gw100} and the related Supplementary Information), slightly better (by a few meV) than those extrapolated to the complete basis-set limit. 


Moving to the discussion of 
$GW$ QP-corrections for the HOMO and LUMO states, the situation becomes more complex.
The $GW$ quasiparticle corrections for HOMO and LUMO levels depend on several numerical parameters.
As detailed in Sec.~\ref{sec:workflow}, we employed an extrapolation scheme with respect to the number of included empty states and the kinetic energy cutoff adopted to represent the response function.
The accuracy of this procedure is analyzed in the \suppinfo.
Overall, calculations done with the tightest convergence parameters are close to full convergence.
This is particularly true for the IP computed within PPA, for which
the Bruneval–Gonze terminator~\cite{BG} (available only for PPA) proved essential for accelerating convergence due to the dense manifolds of empty states in finite systems.


Next, we analyze the $G_0W_0$ IPs obtained with the GN–PPA and MPA schemes of \yambo{}. 
Table~\ref{tab:MAE} summarizes the statistical indicators (ME, MAE, MARE, and $\sigma$) of \yambo{} results with respect to other community codes. 
The GN–PPA data exhibit slightly larger deviations, as expected from the simpler frequency treatment of the plasmon-pole approximation, while MPA results tend to show smaller $\sigma$, overall indicating improved accuracy. 
When compared with other plane-wave or mixed-basis implementations, such as WEST, VASP, and \textsc{FHI-aims}, both approaches attain comparable accuracy.

The individual IPs of all molecules are reported in Table~\ref{tab:HOMO_RES_1}, together with previously published GW100 results~\cite{gw100}, CCSD~\cite{CCSD}, and experimental data from the NIST database~\cite{NIST_db}.
The best agreement for the \yambo{} GN–PPA results, as measured by the MAE, is obtained with WEST$_\mathrm{sol}$ and WEST$_\mathrm{lin}$, followed by VASP and \textsc{FHI-aims} in the complete-basis-set limit.
This level of consistency is remarkable given the methodological differences among these codes, including the choice of basis sets (plane waves, augmented waves, or localized orbitals), the use of pseudopotentials, the different ways to solve the QP equation (full or linearized), and the different frequency-integration schemes. 
Specifically, the frequency integration is performed using AC (\textsc{VASP}), CD (\textsc{WEST}), and MPA (\yambo). 
Overall, these differences have only a minor effect on the computed QP energies, confirming the robustness of modern $G_0W_0$ implementations.

In contrast, larger discrepancies arise when comparing the \yambo{} GN–PPA results with those obtained using the HL-PPA model of \textsc{BGW}. 
Despite both adopting the linearized QP equation, the two models differ in their parametrization of the dielectric response: the HL and GN versions of PPA can yield significantly different QP corrections depending on the system and the treatment of the so-called unfulfilled modes~\cite{stankovski_g_2011,Miglio2012EPJB,Larson2013PRB,rangel_repr_2020}. A semi-analytical comparison between the two indeed shows that the HL–PPA may systematically overestimate QP shifts with respect to GN-PPA~\cite{Valido_2021}.
Conversely, the FF \textsc{BGW} data show a noticeably better agreement with \yambo{}, although the limited number of reported values (19 molecules) prevents a comprehensive comparison.

The \yambo-MPA results generally improve the agreement with all other codes relative to GN–PPA, as reflected by the smaller MAE and MARE values in Table~\ref{tab:MAE}.
The improvement is particularly significant when comparing with WEST$_\mathrm{lin}$ and VASP, confirming that MPA provides an efficient description of frequency dependence, with an accuracy comparable to that of FF-RA, CD, and AC approaches~\cite{Valido_2021,leon2023efficient,Leon2025PhysRevB}. 
The only notable exception is the comparison with \textsc{FHI-aims} results, based on localized def2-QZVP basis sets, prior to complete-basis-set extrapolation, for which slightly larger deviations persist.
It is worth noting that, despite the absence of the Bruneval–Gonze terminator~\cite{BG} in the MPA implementation, which, at the PPA level, has been shown to accelerate convergence, the overall quality of the extrapolated MPA data remains unaffected. 
Finally, the relatively large MAE observed between the \yambo{} GW data and experimental IPs can be attributed to several factors beyond the computational framework, including the neglect of finite-temperature effects and coupling with ionic degrees of freedom, as well as higher-order corrections such as self-consistency and vertex corrections~\cite{gw100}.


\begin{figure*}[]
    \centering
    \includegraphics[width=\textwidth]{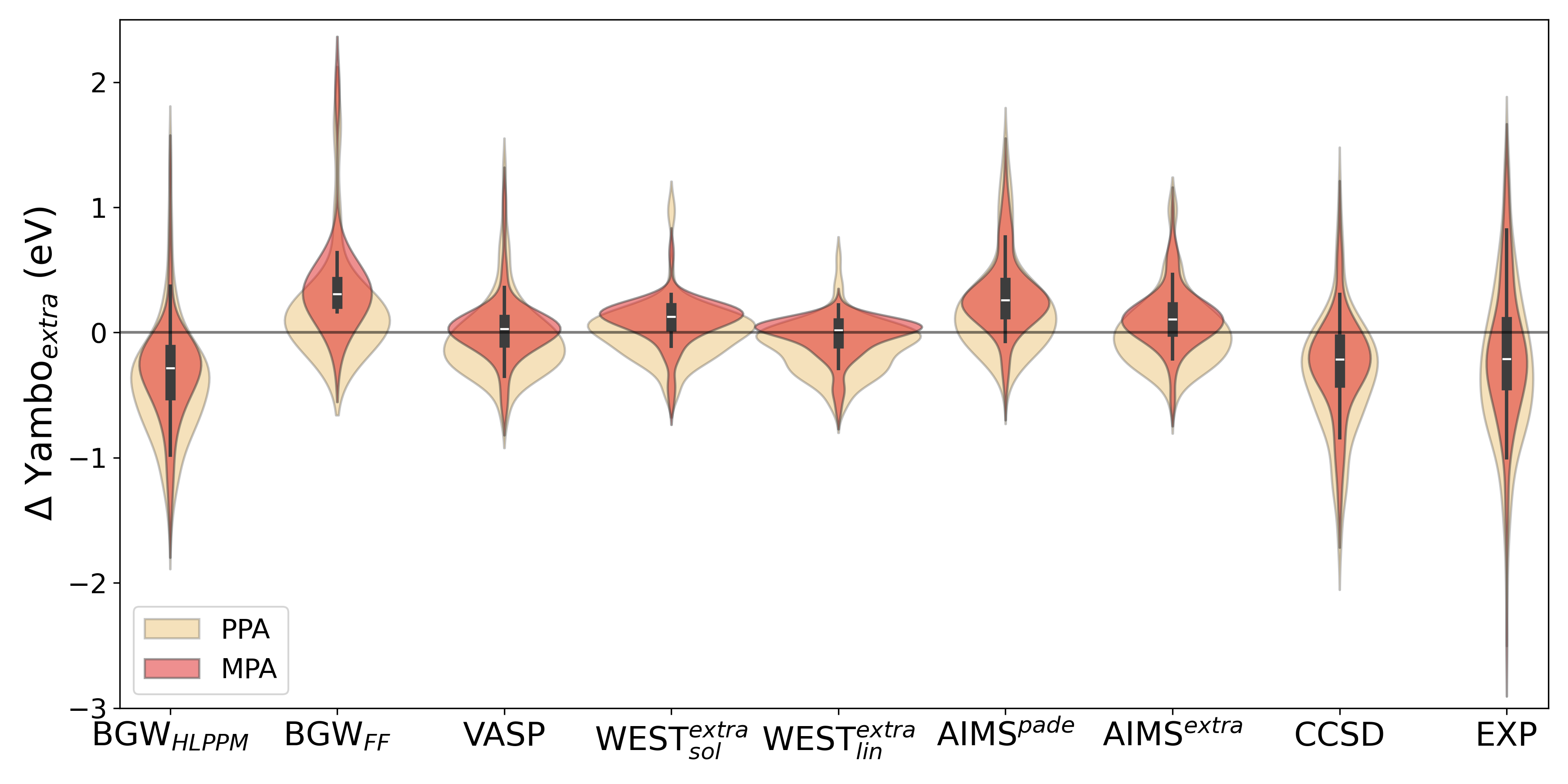}
    \caption{Violin plots representing the distribution of the IP deviation between \yambo{} PPA and MPA results and other GW codes. 
    The width of each curve reflects the density of data points. 
    In the MPA case, the vertical black box represents the inter-quartile range, and the white horizontal line indicates the median.}
    \label{fig:violin}
\end{figure*}

%
%

A more detailed assessment of the deviations across different implementations can be obtained from the violin plots in Fig.~\ref{fig:violin}, which combine box- and density-plot representations. 
The vertical asymmetry of each distribution indicates whether the IPs computed with \yambo{} are systematically over- or under-estimated relatively to the reference data, while the width of the main peak reflects the standard deviation $\sigma$ reported in Tab.~\ref{tab:MAE}.
The values plotted in Fig.~\ref{fig:violin} are summarized in Tab.~\ref{tab:HOMO_RES_1}, with PPA (yellow) and MPA (red) results shown side by side. 
For VASP, 16\% (36\%) of the molecules deviate by approximately $-95$ ($30$)~meV for PPA (MPA), forming the largest peak in each respective distribution. 
When comparing \yambo{}–PPA with WEST$_\mathrm{lin}$, 19\% of the molecules cluster around $-80$~meV, while 32\% form a shoulder at about $-300$~meV. 
In contrast, the MPA results show a narrower and more symmetric distribution, with 35\% of the molecules deviating only by $19$~meV from West$_{lin}$, without any shoulder. 
Overall, \yambo{}–MPA consistently improves the agreement with all other reference codes, further demonstrating the reliability of the MPA formulation.

Most distributions in Fig.~\ref{fig:violin} are asymmetric towards a slight overestimation of the IPs computed with \yambo{} relative to the other methods. 
The exceptions are the comparisons with BGW$_\text{HL-PPA}$, CCSD, and experimental data, where the deviation sign reverses.
Outliers are present for almost all distributions. 
The largest deviation from VASP corresponds to the Xe molecule, with differences of 1.30~eV (PPA) and 1.12~eV (MPA).  
For WEST$_\mathrm{lin}$, the largest outlier is the F$_2$ molecule, with a PPA deviation of 633~meV, while other fluorine-containing molecules such as CF$_4$, FH, and AlF$_3$ show absolute deviations of 400-500~meV. 
These discrepancies are probably due to the presence of strongly localized $2p$ and $3d$ electrons, which are challenging to treat accurately in standard one-shot $G_0W_0$ calculations~\cite{Louie_ZnO}, particularly within PPA \cite{stankovski_g_2011}.
Importantly, the MPA approximation substantially mitigates these discrepancies, reducing, for example, the F$_2$ difference to 273~meV.
The improved performance of the MPA approach is further supported by the cumulative statistics: the number of molecules for which the difference between \yambo{} and VASP (WEST$_\mathrm{lin}$) is smaller than 150~meV increases from 42 (50) in PPA to 72 (73) in MPA.

%
%
Finally, we briefly discuss the main factors influencing the residual discrepancies observed in the present dataset: ($i$) the reduced finite size of the simulation cell, and ($ii$) the linearization of the quasiparticle equation.
Regarding the supercell size, as discussed in Sec.~II of the \suppinfo, the use of a 13 \AA{} FCC cell -- smaller than the 25 \AA{} cubic cells employed in WEST \cite{gov+west+gw100} -- introduces a finite-size error in the $G_0W_0$ ionization potentials. For molecules of average size, this amounts to roughly 25 meV, while for the largest molecules, it increases to about 180~meV.
Consistently, when comparing with codes that use similar computational setups (e.g., WEST$_\mathrm{lin}$ or VASP), the MAEs lie within the expected uncertainty range.

Regarding the linearization of the QP equation, it shows a moderate impact when compared with WEST data obtained with and without linearization, changing the MAEs relative to \yambo{}–PPA and MPA by about 20 and 60~meV, respectively. 
Finally, we did not find any significant correlation between the residual extrapolation error (with respect to the most converged \yambo{} results) and the deviations from other codes, as shown in Fig.~S.5 of the \suppinfo. 
This indicates that differences in the final IPs are only marginally influenced by the extrapolation procedure.

\subsection{Electron Affinities}
\label{sec:EA}
%

%
\begin{table}[]
\centering
\begin{adjustbox}{max width=\linewidth}
\begin{tabular}{cl|cccc}
\hline\hline
    &\\[-5pt]
index & Formula & DFT & \yambo{}$_\mathrm{MPA}$ & \yambo{}$_\mathrm{PPA}$ & WEST$_\mathrm{lin}$\\
& \\[-7pt]
\hline\\[-7pt]
7 & Li$_2$ & 1.75 & 0.73 & 0.76 & 0.58 \\
8 & Na$_2$ & 1.7 & 0.73 & 0.75 & 0.58 \\
9 & Na$_4$ & 1.8 & 1.0 & 1.03 & 1.04 \\
10 & Na$_6$ & 1.43 & 0.87 & 0.93 & 1.0 \\
11 & K$_2$ & 1.33 & 0.72 & 0.79 & 0.72 \\
12 & Rb$_2$ & 1.17 & 0.59 & 0.65 & 0.71 \\
14 & P$_2$ & 3.42 & 1.16 & 1.08 & 1.08 \\
15 & As$_2$ & 3.39 & 1.17 & 1.08 & 1.08 \\
16 & F$_2$ & 5.94 & 0.7 & 0.32 & 1.05 \\
17 & Cl$_2$ & 4.22 & 1.5 & 1.25 & 1.37 \\
18 & Br$_2$ & 4.49 & 2.0 & 1.79 & 1.87 \\
19 & I$_2$ & 4.44 & 3.15 & 3.18 & 3.21 \\
26 & C$_4$ & 6.05 & 3.18 & 2.93 & 3.11 \\
29 & C$_8$H$_8$ & 2.29 & 0.22 & -0.04 & 0.05 \\
35 & CCl$_4$ & 2.69 & 0.55 & 0.31 & 0.39 \\
36 & CBr$_4$ & 3.48 & 1.62 & 1.42 & 1.44 \\
37 & CI$_4$ & 4.11 & 3.1 & 3.06 & 3.03 \\
41 & Si$_5$H$_{12}$ & 1.22 & -0.12 & -0.17 & 0.05 \\
42 & LiH & 1.59 & 0.1 & 0.1 & 0.05 \\
43 & KH & 1.6 & 0.32 & 0.3 & 0.22 \\
54 & F$_2$Mg & 2.64 & 0.31 & 0.25 & 0.32 \\
55 & TiF$_4$ & 4.07 & 0.7 & 0.29 & 0.82 \\
56 & AlF$_3$ & 2.61 & 0.12 & 0.03 & 0.14 \\
58 & SF$_4$ & 3.57 & 3.13 & 3.35 & 0.05 \\
59 & BrK & 1.82 & 0.45 & 0.42 & 0.39 \\
60 & GaCl & 2.39 & 0.46 & 0.45 & 0.42 \\
61 & NaCl & 2.22 & 0.49 & 0.47 & 0.45 \\
62 & MgCl$_2$ & 2.57 & 0.76 & 0.72 & 0.68 \\
63 & AlI$_3$ & 2.58 & 1.57 & 1.54 & 1.65 \\
66 & PN & 3.41 & 0.58 & 0.35 & 0.49 \\
77 & CS$_2$ & 2.79 & 0.56 & 0.33 & 0.48 \\
81 & O$_3$ & 6.17 & 2.56 & 1.87 & 2.6 \\
82 & SO$_2$ & 4.4 & 1.37 & 1.02 & 1.36 \\
83 & BeO & 4.84 & 2.25 & 2.27 & 2.23 \\
84 & MgO & 4.29 & 1.95 & 1.71 & 2.03 \\
93 & C$_5$H$_6$N$_2$O$_2$ & 2.24 & 0.19 & -0.11 & 0.08 \\
94 & C$_4$H$_4$N$_2$O$_2$ & 2.44 & 0.2 & -0.05 & 0.13 \\
96 & Ag$_2$ & 3.08 & 1.54 & 1.52 & 1.47 \\
97 & Cu$_2$ & 3.09 & 1.3 & 1.3 & 1.29 \\
98 & NCCu & 4.12 & 1.91 & 1.85 & 1.92 \\ \\
\hline\hline
\end{tabular}
\end{adjustbox}
\caption{Quasiparticle Electron Affinity for selected molecule of the GW100 dataset as obtained within this work, compared with the available results of West$_{lin}^{extra}$~\cite{gov+west+gw100}. Both DFT and QP values are indicated.}
\label{tab:LUMO_RES}
\end{table}
\begin{figure}
    \centering
    \includegraphics[width=0.45\textwidth]{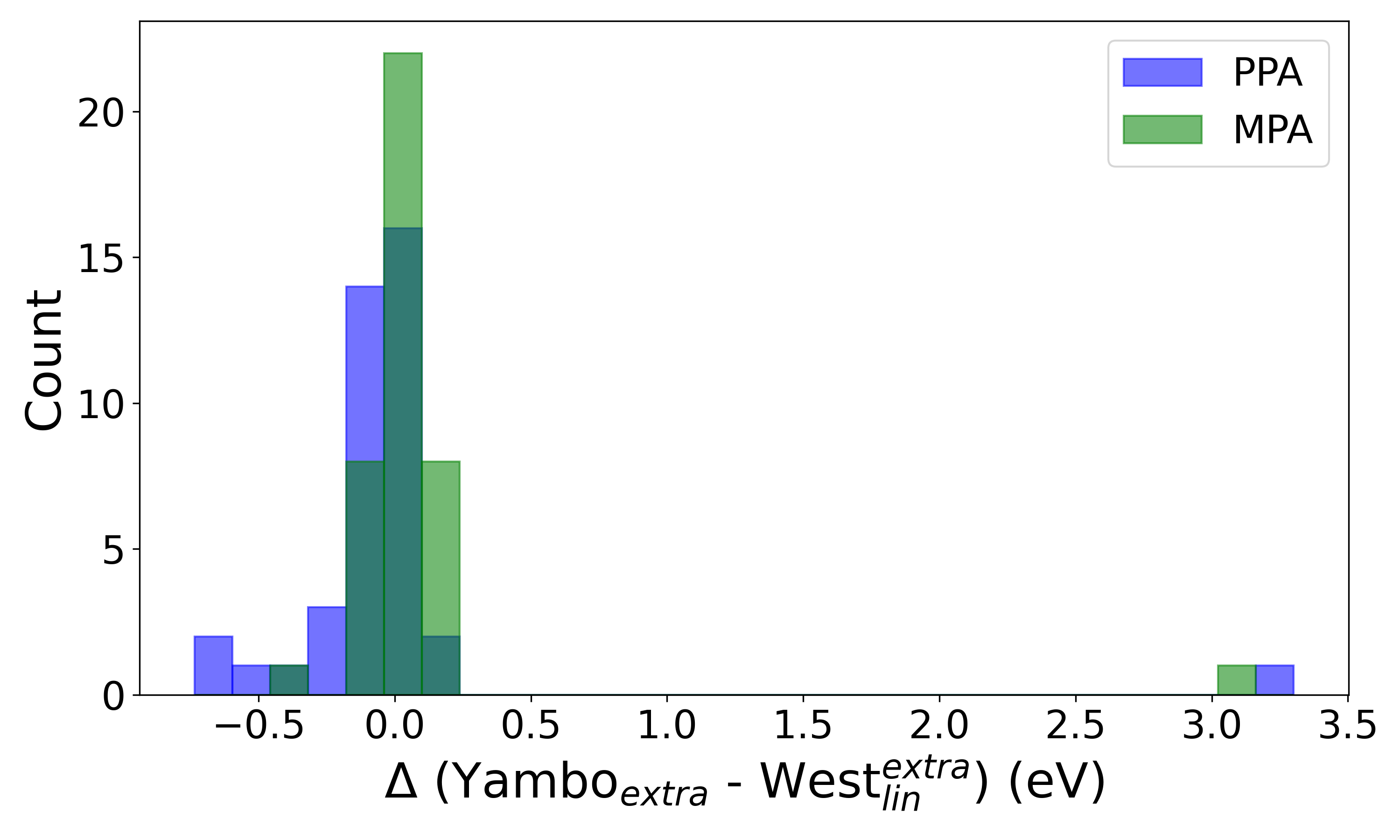}
    \caption{Deviation of the quasiparticle electron affinity (EA) between the \yambo{} and the WEST linearized extrapolated results.}
    
    \label{fig:LUMO_Y_VS_W}
\end{figure}

The analysis of QP EA energies follows the same protocol as for the ionization potentials. Here, the results are compared with those obtained using the WEST code and the linearized $G_0W_0$ QP solution. A summary of the computed values is reported in Tab.~\ref{tab:LUMO_RES}, while Fig.~\ref{fig:LUMO_Y_VS_W} displays the statistical differences between \yambo{} and WEST.

As observed for the IPs, MPA systematically improves the agreement with the reference data compared to PPA. 
The MAE amounts to $163$~meV for MPA and $217$~meV for PPA, confirming that the MPA representation yields a more accurate frequency dependence of the self-energy, including for unoccupied states.
The largest deviation in the dataset corresponds to the SF$_4$ molecule, for which the LUMO exhibits an anomalously high discrepancy (approximately $600$~meV) already at the DFT level~\cite{gov+west+gw100}. 
This deviation propagates to the QP correction and thus accounts for the outlier observed in Fig.~\ref{fig:LUMO_Y_VS_W}. 
Overall, the EA dataset reinforces the conclusions drawn from the IP analysis: the MPA approach achieves accuracy comparable to FF methods at a substantially reduced computational cost.

\section{Conclusions}
\label{sec:conclusions}

In this work, we present a comprehensive benchmark of ionization potentials, IPs, and electron affinities, EAs, for the 100 molecules of the GW100 dataset, computed at the $G_0W_0$ level with the \yambo{} code. The study provides the first systematic assessment of the Godby-Needs plasmon-pole (GN-PPA) and the recently introduced multipole (MPA) approximations, within a plane-wave pseudopotential framework. All calculations were performed using GPU-accelerated resources and automated workflows via the \texttt{aiida-yambo} plugin, ensuring reproducibility and high-throughput efficiency.

Overall, our results show very good agreement with reference data from other $GW$ implementations, including VASP, WEST, BGW, and FHI-aims, with mean absolute errors typically below 0.25~eV. The best agreement is found with WEST$_{\text{lin}}$ and VASP, while deviations with localized-basis codes (AIMS, Turbomole) remain moderate. 
Comparison with CCSD and experimental IPs confirms trends previously reported for the GW100 set, with residual differences largely due to the lack of finite-temperature and electron-phonon effects, as well as the absence of self-consistency and vertex corrections in the present one-shot $G_0W_0$ framework.

Importantly, the comparison between different frequency-integration schemes highlights the reliability of GN-PPA and its partial improvement over the Hybertsen-Louie PPA version used in BGW, which shows larger deviations from full-frequency data. 
MPA further enhances the accuracy of the frequency treatment, providing a good agreement with full-frequency approaches (analytic continuation or contour deformation) at a fraction of their computational cost.

Remaining discrepancies can be mainly attributed to the use of norm-conserving pseudopotentials and the finite size of the supercells. 
 
The linearization of the quasiparticle equation contributes only marginally (20-60~meV). No clear correlation was found between the extrapolation error and deviations with respect to other codes, suggesting that most differences stem from intrinsic methodological choices.

In summary, the present benchmark demonstrates that GN-PPA and, in particular, MPA provide accurate and efficient treatments of $G_0W_0$ quasiparticle calculations, enabling verification against more demanding full-frequency methods. These results consolidate the reliability of the \yambo{} code for large-scale molecular $GW$ simulations. Future developments will target the implementation of full-frequency $GW$ and beyond-$GW$ schemes on GPU architectures, further enhancing both the accuracy and scalability of excited-state calculations.

%
\section*{Data availability}
%
The data presented in this work are available on the Materials Cloud Archive \cite{talirz_materials_2020} at the URL: \url{https://doi.org/10.24435/materialscloud:4a-d7} .

%
\section*{Acknowledgments}
%
We are grateful to M.J. van Setten, M. Govoni, and G. Pizzi for inspiring discussions. This work was partly supported by MaX - MAterials design at the eXascale, the European Centre of Excellence, co-funded by the European High Performance Computing Joint Undertaking (JU) and participating countries within the HORIZON-EUROHPC-JU-2021-COE-1 program (grant n.~101093374); ICSC - Centro Nazionale di Ricerca
in High Performance Computing, Big Data and Quantum Computing, funded by the European Union through the Italian Ministry of University and Research under PNRR M4C2I1.4 (grant n.~CN00000013).
Computational time on the Leonardo machine at CINECA was provided by the EuroHPC regular access program. Computational time on the Juwels-Booster cluster (J\"ulich Supercomputing Center) was provided via the PRCOE07 allocation.
D.A.L.'s work is supported by the Research Council of Norway in the MORTY project (grant n.~664350).



\renewcommand{\emph}{\textit}
\bibliographystyle{achemso}
\bibliography{references}

\end{document}


\renewcommand{\thefigure}{S.\arabic{figure}}
\renewcommand{\thetable}{S.\Roman{table}}

%
\title{Benchmarking the plasmon-pole and multipole approximations in the Yambo Code using the GW100 dataset: Supplementary materials}

\author{M. Bonacci}
\affiliation{PSI Center for Scientific Computing, Theory and Data, 5232 Villigen PSI, Switzerland}
\affiliation{
National Centre for Computational Design and Discovery of Novel Materials (MARVEL), 5232 Villigen PSI, Switzerland
}
\affiliation{
 S3 Centre, Istituto Nanoscienze, CNR, 41125 Modena, Italy
}
\author{D. A. Leon}
\affiliation{
 Department of Mechanical Engineering and Technology Management, \\ Norwegian University of Life Sciences, NO-1432 Ås, Norway
}
\author{N. Spallanzani}
\affiliation{
 S3 Centre, Istituto Nanoscienze, CNR, 41125 Modena, Italy
}
\author{E. Molinari}
\affiliation{
 S3 Centre, Istituto Nanoscienze, CNR, 41125 Modena, Italy
}
\affiliation{Dipartimento di Scienze Fisiche, Informatiche e Matematiche, Universit$\grave{a}$ di Modena e Reggio Emilia, I-41125 Modena, Italy}
\author{D. Varsano}
\affiliation{
 S3 Centre, Istituto Nanoscienze, CNR, 41125 Modena, Italy
}
\author{A. Ferretti}
\affiliation{
 S3 Centre, Istituto Nanoscienze, CNR, 41125 Modena, Italy
}
\author{C. Cardoso}
\email[corresponding author:]{claudiamaria.cardosopereira@nano.cnr.it}
\affiliation{
 S3 Centre, Istituto Nanoscienze, CNR, 41125 Modena, Italy
}

\maketitle
\tableofcontents

\section{Computational details: Resource requirements}

In this Section we complement the computational details provided in the main text by discussing the HPC resource required to run the GW100 dataset within our computational setup. These include the memory requirements as well as the time-to-solution.

\begin{figure}[!b]
    \centering
    \includegraphics[width=\columnwidth]{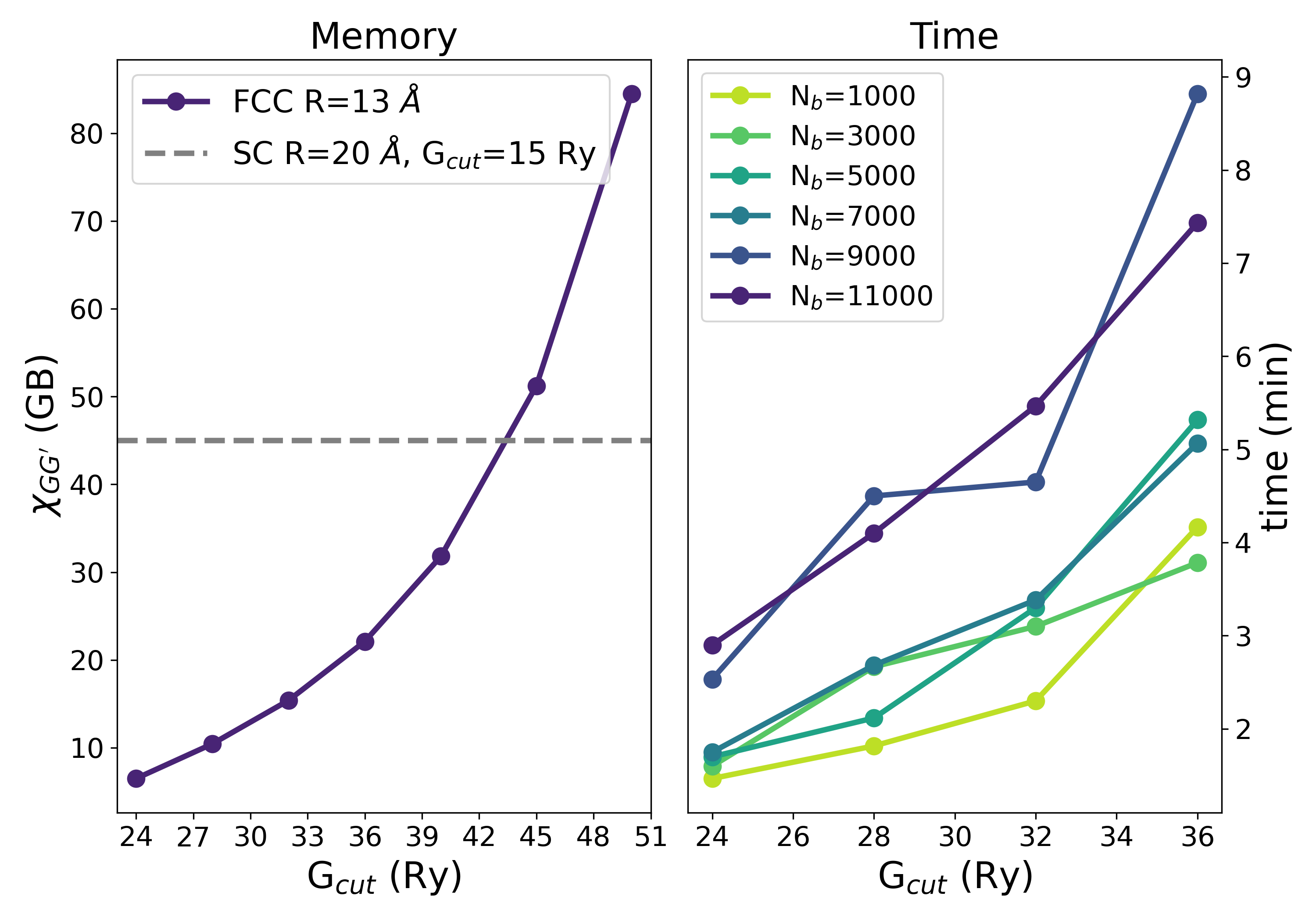}
    \caption{Left panel: memory needed to store the polarizability matrix $\chi_{GG'}$ with respect to the PW cutoff $G_{cut}$. Right panel: time-to-solution with respect to N$_b$ and $G_{cut}$ using GPU acceleration, for the PPA case.}
    \label{fig:size_time}
\end{figure}

Concerning memory requirements, one of the largest pieces of allocated memory of our computational workflow originates from the representation of the response function as a dense $\chi_{\mathbf{G}\mathbf{G}'}$ matrix. As discussed in Sec. II of the main text, this motivated us to adopt an FCC cell rather than a simple-cubic (SC) one,  as the FCC geometry decreases the simulation volume without compromising the minimum distance between periodic images.
In Fig.~\ref{fig:size_time} we show the scaling of the $\chi_{\mathbf{G}\mathbf{G}'}$ matrix size with respect to the plane-wave (PW) cutoff $G_{cut}$, for an FCC cell with 13 \AA\ side. The memory reaches 40~GB for $G_{cut}$ = 43~Ry, while the memory required for G$_{cut}$ = 15~Ry with an SC cell with 20~\AA\ side is already 45~GB, exceeding the capabilities of the GPU used in the present work (NVIDIA A100 with 40 GB RAM.)
One should also consider that this memory limitation is further increased by the fact that we have to store additional quantities such as wavefunctions or temporary workspace for linear algebra operations.
Overall, this memory limitation has been overcome in recent versions of the \yambo{} code by implementing an interface to distributed linear algebra solvers on GPUs (such as cuSOLVERMp from the NVIDIA software stack).

In the following, we discuss the timings and memory requirements for both PPA and MPA calculations.
Automating large numbers of simulations—particularly within the computationally demanding MBPT framework requires the use of high performance computing (HPC). Modern HPC architectures very often take advantage of hardware accelerators based on graphical processing units (GPUs), significantly reducing the time- and energy- to-solution with respect to CPU-only machines. 
In Fig.~\ref{fig:size_time} we show the wall-time for a typical $G_0W_0$ flow (with frequency integration treated at the PPA level)  performed on the Juwels-Booster cluster \cite{JUWELS} (J\"ulich, Germany) using 20-64 nodes, each of them equipped with 4 A100 GPUs with 40 GB of memory.
Each individual calculation takes less than 10 minutes to be completed, leading to a total of 1.5 hours per molecule, when considering 24 simulations. This performance is enabled by the significant speed-up provided by GPU-accelerated machines and codes. 

On the basis of the above analysis, the predicted wall-time required to run the whole GW100 set at the PPA level was then estimated in about 150 hours.

In practice, the wall-time was reduced to $\sim$150/8 h = 18.75 hours, since on average eight calculations were running simultaneously and automatically, i.e. without human intervention to submit new jobs, thanks to the AiiDA scheduling system.
In contrast, MPA calculations were computed using 20–64 nodes of the Leonardo supercomputer (Cineca, Italy), equipped with 4 NVIDIA A100 GPUs (64 GB of memory) per node. Simulations took $\sim$13 hours  per molecule and a total time of 1100 hours (with an effective wall-time of about 140 hours).  We underline that while MPA simulations are generally more computationally expensive than PPA. Moreover, N$_b$ was also set to larger values in the MPA case (we use, on average, 2000 more empty states for each calculation), as the MPA does not employ the terminator technique~\cite{BG}, which accelerates convergence with respect to the summation over empty states.

%
\section{Supercell choice}
\label{supercell}
%
Calculations of low-dimensional systems using plane-wave codes, which implicitly assume periodic boundary conditions, require large unit cells to include sufficient vacuum and avoid spurious interactions between replicas. At fixed cutoff, larger cells lead to  more plane-waves. Consequently, when computing the response function in reciprocal space, the size of the screening matrix scales quadratically with the number of PWs, or equivalently, with the cube of the kinetic energy cutoff $G_{cut}$ and linearly with the volume of the cell (see Fig.~\ref{fig:size_time}). 
For this reason, keeping the supercell as small as possible is highly desirable.
Previous GW100 calculations~\cite{gov+west+gw100} used a simple cubic (SC) supercell with lattice parameters as large as 25 \AA. 
In this work, we employ a face-centered cubic (FCC) cell while keeping the same distances between the nearest neighboring (NN) molecules. This choice reduces the supercell volume by a factor of $\sqrt{2}$ compared to the SC cell. 

\begin{figure}
    \centering
    \includegraphics[width=0.95\columnwidth]{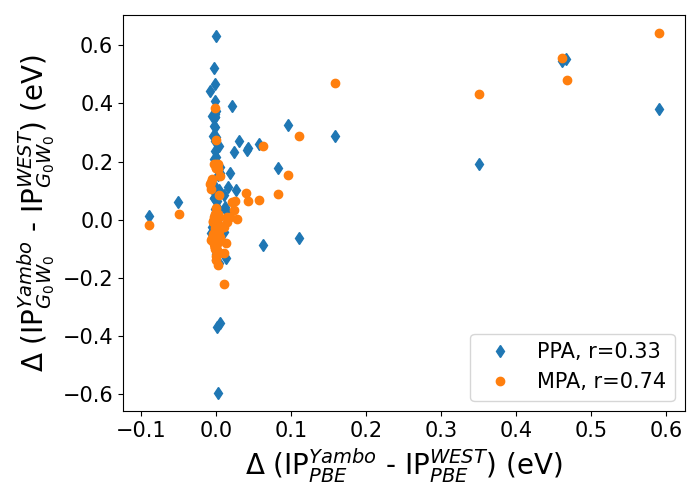}
    \caption{Plot of the deviation between the DFT PBE IP energies obtained in this work and in the WEST GW100 paper~\cite{gov+west+gw100} versus the same deviation but for the $G_0W_0$ IP energies, for both the linearized and the secant solution obtained within the WEST code. In the legend, the correlation \textit{r} between the $G_0W_0$ and DFT deviations is reported.}
    \label{fig:correlation_DFT_G0W0}
\end{figure}

The DFT PBE HOMO energies are shown in Tables \ref{tab:HOMO_DFT_1} and \ref{tab:HOMO_DFT_2}. 
Considering the reduced amount of vacuum used here, the DFT PBE values agree well with previously reported results~\cite{gov+west+gw100}. In particular, the average discrepancy is 30~meV, error smaller than 10~meV for more than 74\% of the molecules. There is a strong positive correlation, r=0.74 (Fig.~\ref{fig:correlation_DFT_G0W0}), in the difference between the HOMO energies obtained with PBE and MPA@G$_0$W$_0$ and those obtained with WEST~\citenum{gov+west+gw100}. 
In contrast, for PPA@G$_0$W$_0$, the correlation is weak ($r=0.33$), consistent with the simpler treatment of the frequency dependence with respect to MPA.
Notably, the two deviation sets plotted in Fig~\ref{fig:correlation_DFT_G0W0} share part of the outliers (for example the ones around 0.5-0.6 eV). 
This indicates that the agreement between the two codes can be improved by using a larger supercell, and that some of the outliers are mostly due to the different DFT starting points. 
To estimate the possible improvement, we performed a statistical analysis similar to the one described in the main text for Table~II, now considering only molecules with a discrepancy between YAMBO and WEST DFT results smaller than 5 meV (i.e. 64\% of the molecules in the set). The results are summarized in Table~\ref{tab:SI_MAE_less_data}, showing that only a few tens of meV improvement is obtained, mainly for the comparison with WEST$^{extra}_{lin}$ and VASP.
This demonstrates that while the DFT-level deviation due to the reduced supercell is measurable, its propagation to the $G_0W_0$ results has only a limited impact on the statistical analysis and does not affect the overall conclusions presented in the main text.

\begin{table*}[]
\begin{tabularx}{\textwidth}{
l|
>{\centering\arraybackslash}X
>{\centering\arraybackslash}X|
>{\centering\arraybackslash}X
>{\centering\arraybackslash}X|
>{\centering\arraybackslash}X
>{\centering\arraybackslash}X|
>{\centering\arraybackslash}X
>{\centering\arraybackslash}X}
\hline\hline
    &&&&&&&&\\[-7pt]
\textbf{Code} & \multicolumn{2}{c|}{\textbf{ME} (meV)}   & \multicolumn{2}{c|}{\textbf{MAE} (meV)}   & \multicolumn{2}{c|}{\textbf{MARE} ($\%$)} & \multicolumn{2}{c}{ $\boldsymbol{\sigma}$ (meV)}\\[4pt]
    & GN-PPA & MPA &GN-PPA & MPA &GN-PPA & MPA &GN-PPA & MPA \\[4pt]
    \hline
    &&&&&&&&\\[-7pt]
    YAMBO$_\text{GN-PPA}$ & - & 103 & - & 168 & - & 1.6 & - & 116 \\[4pt]
WEST$^{extra}_{lin}$ & -91 & 11 & 166 & 86 & 1.5 & 0.8 & 164 & 68 \\[4pt]
WEST$^{extra}_{sol}$ & 57 & 159 & 159 & 174 & 1.7 & 1.7 & 187 & 109 \\[4pt]
VASP & -22 & 80 & 210 & 126 & 2.0 & 1.2 & 209 & 188 \\[4pt]
AIMS$^{pade}$ & 239 & 342 & 308 & 351 & 3.3 & 3.5 & 307 & 244 \\[4pt]
AIMS$^{extra}$ & 59 & 174 & 213 & 196 & 2.2 & 2.0 & 224 & 186 \\[4pt]
BGW$_\text{HL-PPA}$ & -438 & -335 & 516 & 395 & 4.2 & 3.3 & 364 & 331 \\[4pt]
BGW$_\text{FF}$ & 200 & 394 & 264 & 407 & 2.7 & 4.0 & 393 & 398 \\[4pt]
CCSD & -381 & -273 & 459 & 355 & 3.8 & 3.0 & 374 & 317 \\[4pt]
EXP & -249 & -142 & 525 & 456 & 4.4 & 3.9 & 449 & 436 \\[4pt]
\hline\hline
\end{tabularx}
\caption{Statistical errors between Yambo and other codes and experiments considering only a reduced set of the GW100 molecules, where the deviation between the DFT HOMO energy computed within the reduced supercell and the one of Ref.~\citenum{gov+west+gw100} is less than 5 meV.
}
\label{tab:SI_MAE_less_data}
\end{table*}

%

%
%
\begin{figure}
    \centering
    \includegraphics[width=\columnwidth]{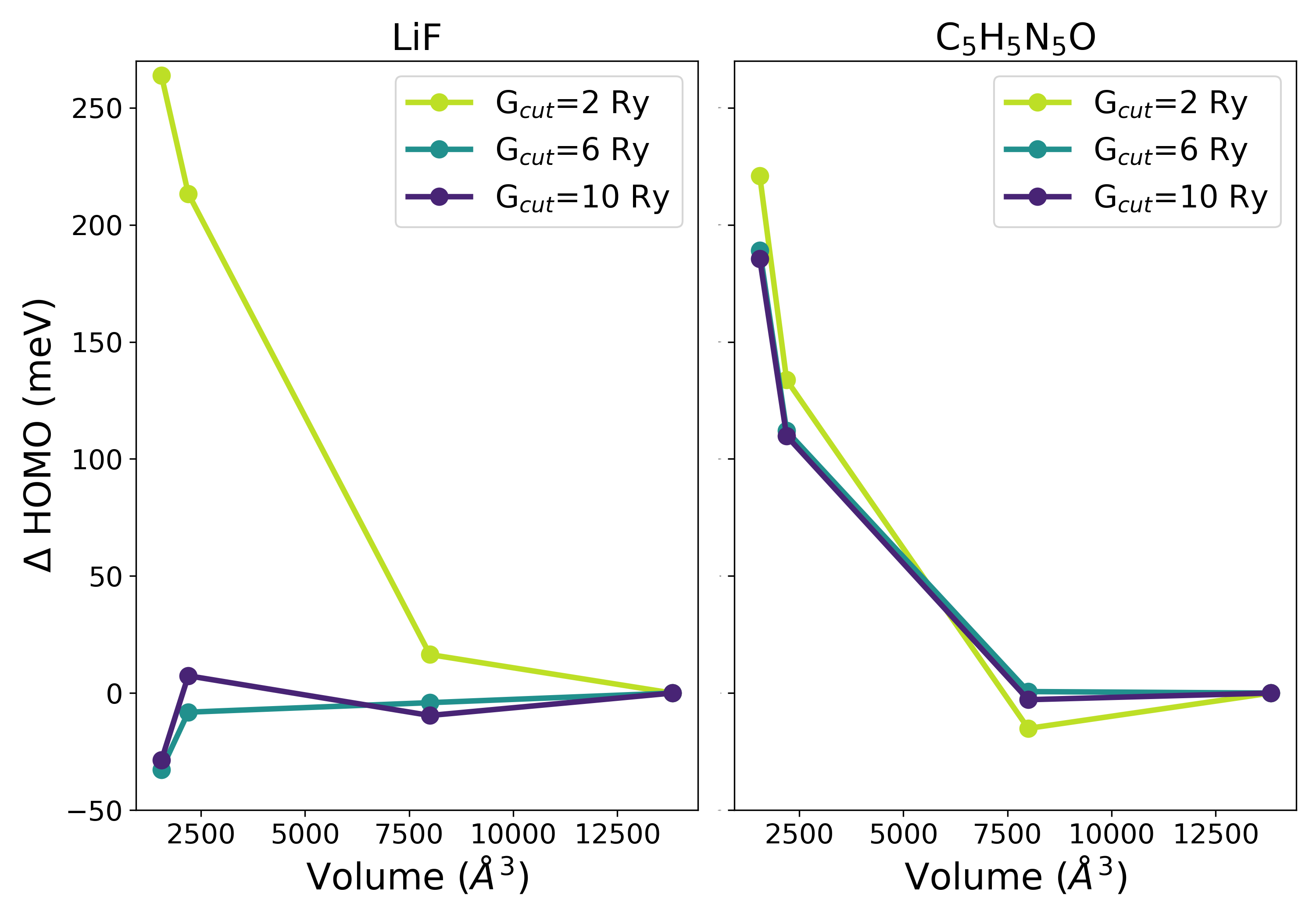}
    \caption{Convergence of the (PPA) HOMO level for two molecules, LiF and C$_5$H$_5$N$_5$O. The number of bands is kept fixed, N$_b$=0.75 Ry, while varying $G_{cut}$ and the cell volume. The first point corresponds to an FCC cell with lattice \textbf{R}=13 \AA. The other 3 points correspond to SC cells of lattice vector \textbf{R}=13, 20, and 24 \AA. The results obtained for the maximum cell volume are set to zero for each fixed value of $G_{cut}$.
    }
    \label{fig:vacuum_conv}
\end{figure}

To provide a more direct estimate of the error in the quasiparticle calculations, we performed convergence tests with respect to the supercell volume for two representative molecules: LiF and C$_5$H$_5$N$_5$O (see Fig.~\ref{fig:vacuum_conv}).
The first has 10, while the latter has 56 valence electrons and the largest linear size in the GW100 set, 7.5~\AA.
We considered a 13~\AA\ FCC cell and three cubic cells with side lenghts 13, 20, and 24~\AA\ while using underconverged N$_b$ and $G_{cut}$ values. The results are shown in Fig.~\ref{fig:vacuum_conv}.
DFT calculations are performed using the Martyna-Tuckerman method~\cite{martyna_t}, which cures spurious electrostatic effects due to the long-range Coulomb interaction. 
Instead, the $GW$ calculations use a spherical truncation of the Coulomb potential with a diameter $\sim1$\AA\ shorter than the lattice parameter. Notably, this Coulomb cutoff analytically removes the divergence of the long-range Coulomb interaction $v(\mathbf{q}\rightarrow0)$~\cite{rozzi_exact_2006}.
Numerically, we find that the results are converged within $\sim$25 and $\sim$180~meV for LiF and C$_5$H$_5$N$_5$O, respectively. Considering that all other molecules are smaller than C$_5$H$_5$N$_5$O and that the final results are computed with larger convergence parameters, we estimate the error due to the use of a finite FCC cell to have an upper bound of 180~meV. 

\section{The extrapolation procedure}

\begin{figure}[H]
    \centering
    \includegraphics[width=\columnwidth]{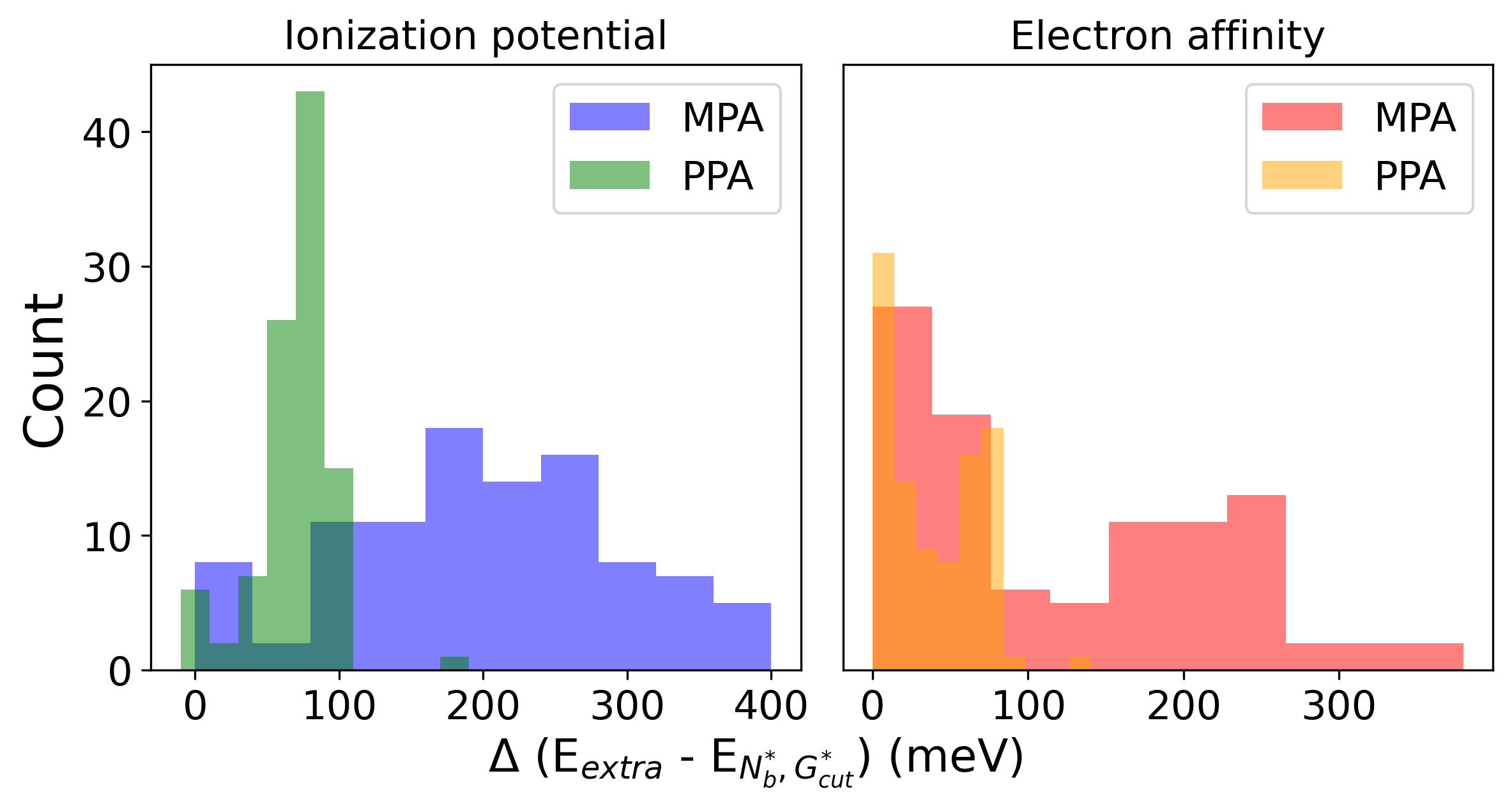}
    \caption{Convergence evaluation of \yambo{} results for IP (left panel) and EA (right panel) with respect to the extrapolated value.}
    \label{fig:conv}
\end{figure}

%
We start by comparing IP and EA computed with the tightest convergence parameters with the corresponding extrapolated values, according to the procedure described in Sec. II of the main text. 
In Fig.~\ref{fig:conv} we show an histogram of frequency distribution of the deviation between the two set of results, and in Tab.~\ref{tab:MAE_extrap_explicit} we present the mean absolute error (MAE) and the standard deviation ($\sigma$) of the same data. For PPA, we have a MAE ($\sigma$) of 68 (25) and 38 (30) meV, respectively, for IP and EA. MPA shows considerably larger values of both MAE and $\sigma$, showing that the calculated results are further away from the extrapolated values. The MAE ($\sigma$) values for IP and EA are, respectively, 199 (95) and 121 (99) meV. This can be  explained by the fact that MPA, contrary to PPA, does not take advantage of the terminator technique~\cite{BG} that accelerates the convergence with respect to the summation over empty states. Notably, these deviations are still rather small since the mean absolute relative error (MARE) for the IP is only 0.65$\%$ (1.8$\%$) for PPA (MPA). 

We performed the same analysis regarding the evaluation of the HOMO-LUMO quasiparticle gap (Tab.~\ref{tab:MAE_extrap_explicit}), which is expected to converge faster than the individual states. In fact, the MAE significantly decreases. 

\begin{table}
\begin{tabularx}{\columnwidth}{
l|
>{\centering\arraybackslash}X
>{\centering\arraybackslash}X|
>{\centering\arraybackslash}X
>{\centering\arraybackslash}X}
\hline\hline
    &&&&\\ [-7pt]
     & \multicolumn{2}{c|}{MAE (meV)}  & \multicolumn{2}{c}{$\sigma$ (meV)} \\ 
                & PPA & MPA & PPA & MPA \\ [4pt]
    \hline
    &&&&\\ [-7pt]
    IP   & 68  & 199  & 25  & 95 \\ 
    EA   & 38 & 121  & 30 & 99 \\ 
    IP-EA  & 29  & 104  & 80 & 95 \\[4pt]
    \hline\hline
\end{tabularx}
\caption{Mean absolute error (MAE), and standard deviation ($\sigma$) for the deviation from extrapolated and most converged quasiparticle HOMO-LUMO gaps.}
\label{tab:MAE_extrap_explicit}
\end{table}

\begin{figure*}
    \centering
    \includegraphics[width=0.9\textwidth]{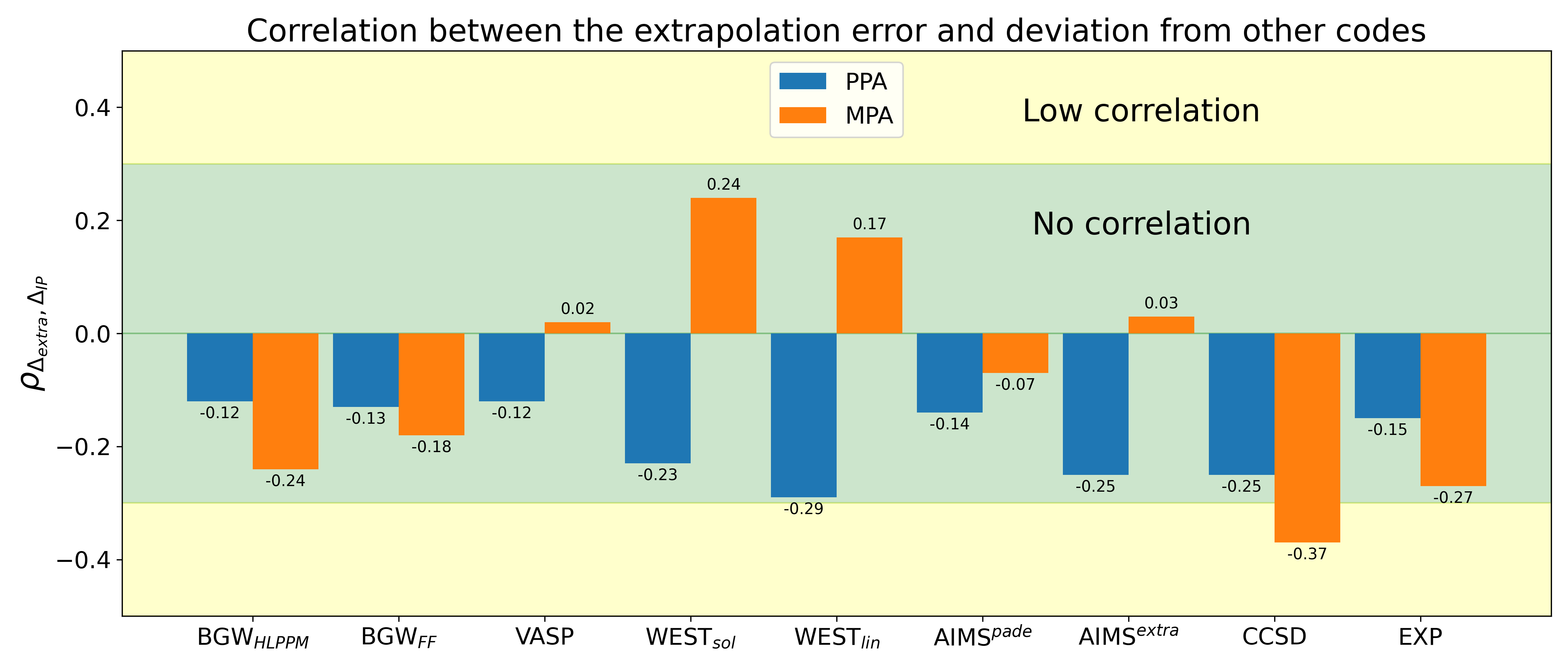}
    \caption{Correlation between the error given by the extrapolation procedure ($\Delta_{extra}$) and the discrepancies with other codes ($\Delta_{IP}$), indicated as $\rho_{\Delta_{extra},\Delta_{IP}}$. The correlation is considered to be significant only if larger than 0.5 (which is never the case for these results).}
    \label{fig:conv_valid}
\end{figure*}

As an additional validation test for the extrapolation procedure, we computed the correlation between the error given by the extrapolation (Fig. \ref{fig:conv}) and the discrepancy between Yambo results and other codes. Results, shown in Fig. \ref{fig:conv_valid}, indicate that there is no correlation between these two quantities, i.e. the extrapolation adds no relevant noise to the overall comparison.

\begin{table*}[]
\centering
\begin{tabular}{cl|c}
\hline\hline
    &\\[-5pt]
index & Formula & HOMO (PBE)\\
& \\[-7pt]
\hline\hline\\[-9pt]
1 & He & -15.76 \\
2 & Ne & -13.33 \\
3 & Ar & -10.26 \\
4 & Kr & -9.25 \\
5 & Xe & -8.23 \\
6 & H$_2$ & -10.38 \\
7 & Li$_2$ & -3.23 \\
8 & Na$_2$ & -3.12 \\
9 & Na$_4$ & -2.33 \\
10 & Na$_6$ & -2.4 \\
11 & K$_2$ & -2.51 \\
12 & Rb$_2$ & -2.37 \\
13 & N$_2$ & -10.29 \\
14 & P$_2$ & -7.11 \\
15 & As$_2$ & -6.49 \\
16 & F$_2$ & -9.42 \\
17 & Cl$_2$ & -7.28 \\
18 & Br$_2$ & -6.8 \\
19 & I$_2$ & -6.26 \\
20 & CH$_4$ & -9.46 \\
21 & C$_2$H$_6$ & -8.16 \\
22 & C$_3$H$_8$ & -7.76 \\
23 & C$_4$H$_10$ & -7.57 \\
24 & C$_2$H$_4$ & -6.77 \\
25 & C$_2$H$_2$ & -7.19 \\
26 & C$_4$ & -7.26 \\
27 & C$_3$H$_6$ & -7.06 \\
28 & C$_6$H$_6$ & -6.33 \\
29 & C$_8$H$_8$ & -5.28 \\
30 & C$_5$H$_6$ & -5.4 \\
31 & C$_2$H$_3$F & -6.55 \\
32 & C$_2$H$_3$Cl & -6.43 \\
33 & C$_2$H$_3$Br & -5.83 \\
34 & C$_2$H$_3$I & -6.04 \\
35 & CF$_4$ & -10.43 \\
36 & CCl$_4$ & -7.66 \\
37 & CBr$_4$ & -6.92 \\
38 & CI$_4$ & -6.13 \\
39 & SiH$_4$ & -8.51 \\
40 & GeH$_4$ & -8.36 \\
41 & Si$_2$H$_6$ & -7.27 \\
42 & Si$_5$H$_12$ & -6.1 \\
43 & LiH & -4.36 \\
44 & KH & -3.48 \\
45 & BH$_3$ & -8.5 \\
46 & B$_2$H$_6$ & -7.88 \\
47 & NH$_3$ & -6.16 \\
48 & HN$_3$ & -6.82 \\
49 & PH$_3$ & -6.72 \\
50 & AsH$_3$ & -6.78 \\ \\
    \hline\hline
\end{tabular}
\caption{DFT PBE HOMO results as computed in this work.}
\label{tab:HOMO_DFT_1}
\end{table*}
\begin{table*}[]
\centering
\begin{tabular}{cl|c}
\hline\hline
    &\\[-5pt]
index & Formula & IP PBE\\
& \\[-7pt]
\hline\hline\\[-9pt]
51 & SH$_2$ & -6.29 \\
52 & FH & -9.65 \\
53 & ClH & -8.03 \\
54 & LiF & -6.13 \\
55 & F$_2$Mg & -8.31 \\
56 & TiF$_4$ & -10.45 \\
57 & AlF$_3$ & -9.72 \\
58 & BF & -6.78 \\
59 & SF$_4$ & -8.37 \\
60 & BrK & -4.72 \\
61 & GaCl & -6.58 \\
62 & NaCl & -5.29 \\
63 & MgCl$_2$ & -7.61 \\
64 & AlI$_3$ & -6.48 \\
65 & BN & -7.47 \\
66 & NCH & -9.04 \\
67 & PN & -7.74 \\
68 & H$_2$NNH$_2$ & -5.28 \\
69 & H$_2$CO & -6.27 \\
70 & CH$_4$O & -6.34 \\
71 & C$_2$H$_6$O & -6.16 \\
72 & C$_2$H$_4$O & -5.97 \\
73 & C$_4$H$_10$O & -5.77 \\
74 & CH$_2$O$_2$ & -6.95 \\
75 & HOOH & -6.45 \\
76 & H$_2$O & -7.25 \\
77 & CO$_2$ & -9.1 \\
78 & CS$_2$ & -6.79 \\
79 & OCS & -7.48 \\
80 & OCSe & -6.94 \\
81 & CO & -9.34 \\
82 & O$_3$ & -7.95 \\
83 & SO$_2$ & -8.04 \\
84 & BeO & -6.16 \\
85 & MgO & -4.82 \\
86 & C$_7$H$_8$ & -5.97 \\
87 & C$_8$H$_10$ & -5.92 \\
88 & C$_6$F$_6$ & -6.64 \\
89 & C$_6$H$_5$OH & -5.62 \\
90 & C$_6$H$_5$NH$_2$ & -5.0 \\
91 & C$_5$H$_5$N & -5.92 \\
92 & C$_5$H$_5$N$_5$O & -5.21 \\
93 & C$_5$H$_5$N$_5$ & -5.49 \\
94 & C$_4$H$_5$N$_3$O & -5.71 \\
95 & C$_5$H$_6$N$_2$O$_2$ & -6.0 \\
96 & C$_4$H$_4$N$_2$O$_2$ & -6.27 \\
97 & CH$_4$N$_2$O & -5.93 \\
98 & Ag$_2$ & -5.19 \\
99 & Cu$_2$ & -4.74 \\
100 & NCCu & -6.78 \\ \\
    \hline\hline
\end{tabular}
\caption{Continuation of Table \ref{tab:HOMO_DFT_1}.}
\label{tab:HOMO_DFT_2}
\end{table*}
%

%
%
\renewcommand{\emph}{\textit}
\bibliographystyle{achemso}
\bibliography{references}